%% file: main.tex
\PassOptionsToPackage{prologue,dvipsnames}{xcolor}

\documentclass[acmsmall,screen]{acmart}

\AtBeginDocument{%
  }

\setcopyright{acmcopyright}
\copyrightyear{2018}
\acmYear{2018}
\acmDOI{XXXXXXX.XXXXXXX}

\acmConference[]{}{}{}
\acmPrice{15.00}
\acmISBN{978-1-4503-XXXX-X/18/06}

\usepackage{multirow}
\usepackage{graphicx}
\usepackage{diagbox}
\usepackage{xspace}
\usepackage[ruled, vlined, linesnumbered]{algorithm2e}
\usepackage{bbding}
\usepackage{threeparttable}
\usepackage{amsmath}
\usepackage{adjustbox}
\usepackage{listings}
\usepackage{lstlinebgrd} % The add-on for line backgrounds
\usepackage[T1]{fontenc}
\usepackage{semantic}
\usepackage{newfloat}
\usepackage{enumitem}
\usepackage{booktabs}
\usepackage{arydshln}
\usepackage{makecell}
\usepackage{subcaption}
\usepackage{tcolorbox}

\usepackage[scaled=0.85]{beramono}

\usepackage{framed}
\usepackage{hyperref}
\hypersetup{hidelinks,
	colorlinks=true,
	allcolors=black,
	pdfstartview=Fit,
	breaklinks=true}

\usepackage{pifont}% http://ctan.org/pkg/pifont
\newcommand{\CodeIn}[1]{{\small\texttt{#1}}}

\DeclareFloatingEnvironment[fileext=frm,placement={!ht},name=Rules]{infrule}

\lstdefinestyle{cot}{
    basicstyle = \ttfamily\scriptsize,           % 基本样式 + 小号字体
    breaklines = true,                  % 代码过长则换行
    breakindent=0pt,
    keywordstyle = \color{blue},            % 关键字颜色
    stringstyle = \color{red!100},          % 字符串颜色
    frame = shadowbox,                  % 用（带影子效果）方框框住代码块
    escapeinside={(*@}{@*)},
}

\lstdefinestyle{code}{
    basicstyle = \ttfamily\scriptsize,           % 基本样式 + 小号字体
    breaklines = true,                  % 代码过长则换行
    columns=fullflexible,
    breakindent=0pt,
    keywordstyle = \color{blue},            % 关键字颜色
    stringstyle = \color{red!100},          % 字符串颜色
    showspaces = false,                 % 不显示空格
    showstringspaces=false,
    escapeinside={(*@}{@*)},
}

\newcommand\dataset{Repo-Smith-Test}
\newcommand\approach{Repo-Smith}

\definecolor{deep-blue}{RGB}{0,0,200}
\definecolor{deep-red}{RGB}{200,0,0}

\begin{document}

\title{Synthesizing File-Level Data for Unit Test Generation with Chain-of-Thoughts via Self-Debugging}


% \author{Anonymous}
% \affiliation{%

% \affiliation{%

\author{Ziyue Hua}
\affiliation{%
  \institution{Key Lab of HCST (PKU), MOE; SCS; Peking University}
  \country{China}
}
\email{ziyue_hua@stu.pku.edu.cn}

\author{Tianyu Chen}
\authornote{Corresponding author}
\email{chentianyu@microsoft.com}
\affiliation{%
  \institution{Microsoft Research}
  \country{China}  
}

\author{Yeyun Gong}
\email{yegong@microsoft.com}
\affiliation{%
  \institution{Microsoft Research}
  \country{China}  
}

\author{Shuai Lu}
\email{shuailu@microsoft.com}
\affiliation{%
  \institution{Microsoft Research}
  \country{China}  
}
\author{Peng Cheng}
\email{pengc@microsoft.com}
\affiliation{%
  \institution{Microsoft Research}
  \country{China}  
}

\author{Qinglin Zhu}
\email{qinglin.1.zhu@kcl.ac.uk}
\affiliation{%
  \institution{King's College London}
  \country{UK}
}

\author{Yibo He}
\affiliation{%
  \institution{Key Lab of HCST (PKU), MOE; SCS; Peking University}
  \country{China}
}
\email{yibohe@pku.edu.cn}
\author{Yingjie Fu}
\affiliation{%
  \institution{Key Lab of HCST (PKU), MOE; SCS; Peking University}
  \country{China}
}
\email{yingjiefu@stu.pku.edu.cn}

\author{Wenpin Jiao}
\affiliation{%
  \institution{Key Lab of HCST (PKU), MOE; SCS; Peking University}
  \country{China}
}
\email{jwp@sei.pku.edu.cn}

\author{Wei Yang}
\affiliation{%
  \institution{UT Dallas}
  \country{USA}
}
\email{wei.yang@utdallas.edu}

\author{Tao Xie}
\affiliation{%
  \institution{Key Lab of HCST (PKU), MOE; SCS; Peking University}
  \country{China}
}
\email{taoxie@pku.edu.cn}

% \affiliation{%

% \affiliation{%

% \affiliation{%

% \affiliation{%

% \affiliation{%

\renewcommand{\shortauthors}{Anonymous}

\begin{abstract}
Automatic unit test (UT) generation is essential for software quality assurance, but existing approaches—including symbolic execution, search-based approaches, and recent LLM-based generators—struggle to produce human-quality tests with correct, meaningful assertions and reliable chain-of-thought (CoT) explanations. 
We identify a gap in UT training data: repository-mined tests lack developer CoTs, while LLM-distilled CoTs are often incorrect or incomplete. 
To address this issue, we propose a novel data-distillation approach that uses self-debugging to produce high-quality UT training examples paired with faithful CoTs. 
Our approach combines (1) guided test repair, a heuristic loop (error-, failure-, and coverage-focused steps) that asks the used model to diagnose and iteratively fix generated tests, and (2) CoT compression, which compacts original and debugging CoTs into concise explanations that directly justify correct tests. 
We apply this pipeline to a large corpus of open-source projects to construct a dataset of 74,518 high-quality <focal method, test, CoT> examples, and then use it for supervised fine-tuning of a base model.
An empirical evaluation shows that the fine-tuned model achieves high UT generation effectiveness: it attains a pass rate of 36.17\% on test assertions, a branch coverage of 43.90\%, and a mutation score of 88.66\%, substantially higher than state-of-the-art commercial models like o4-mini. 
% We release the dataset and model artifacts to support future research in reliable, explainable test synthesis.
 \end{abstract}





\maketitle
\input{introduction}

\input{background}

\input{approach}
\input{evaluation}
\input{discuss}
\input{threats}
\input{related}
\input{conclusion}

\input{data_availability}

\bibliographystyle{ACM-Reference-Format}
\bibliography{main}

\end{document}

%% file: introduction.tex
\section{Introduction}~\label{sec: intro}

Automatic unit test (UT) generation plays a critical role in software quality assurance by aiming to efficiently detect software defects through comprehensive coverage of the software under test. 
Traditional approaches, such as symbolic execution~\cite{reanu2008combining, xie2005symstra}, evolutionary algorithms~\cite{evosuite}, and model checking~\cite{gargantini1999using, enoiu2016automated} have been employed to automatically generate unit tests. 
% Although these approaches can achieve promising code coverage, they often fail to match the practical utility of human-written tests, particularly in producing accurate and meaningful assertions.
Despite achieving promising code coverage, these approaches often fail to match the practical utility of human-written tests, particularly regarding accurate and meaningful assertions.
To address this limitation, recent research leverages Large Language Models (LLMs) to automate unit test generation~\cite{cleantest, coderm, hits, yang2024empirical}. 
Note that an LLM's capability in test generation mainly depends on its training data; collecting or synthesizing high-quality unit test datasets substantially improves the LLM's capability in this task.

Existing training data for unit test generation is derived from two principal approaches.
(1) Data collection approaches~\cite{methods2test, pymethods2test, focalstudy}.
% For a specific repository, these approaches collect and filter test cases in this repository with static analysis to determine each corresponding focal method.
These approaches use static analysis to collect and filter test cases in open-source repositories, identify the corresponding focal method for each test case, and result in <focal method, test case> pairs as training data.
The main advantage of these approaches is the correctness and completeness of test cases as these test cases are manually written by developers.
(2) Data distillation approaches~\cite{seed2025seed, phi, phi15}. % yibo: only open-source LLMs? What about LLMs like GPT-4? open-source -> advanced.
These approaches distill advanced LLMs with collected focal methods to generate the targeted test cases for these focal methods.
The main advantage of these approaches is that besides the output test cases, LLMs can generate Chain-of-Thoughts (CoTs), which analyze the given focal method and explain how the corresponding test case is generated step-by-step.
Recent work~\cite{yang2024chain, zhu2025uncertainty} has shown that CoT can substantially improve the LLM's capability in complicated tasks including UT generation.

However, existing approaches for UT generation data face limitations in generating correct and complete cases with CoTs, resulting in the limitation of LLM's capability in high-quality unit test generation.
\textit{Data collection}, collects focal methods and test cases from open-source projects; however, these test cases lack corresponding CoTs. 
Even given the focal methods and the ground-truth test cases, the generated CoTs tend to explain the ground-truth test cases, which do not occur in a real-world scenario of test case generation, thus failing to catch the key thinking points to lead to the test cases.
\textit{Data distillation}, depends on LLMs to generate CoT, but fails to ensure the correctness of the generated CoTs. 
Existing work has shown that LLM-generated test cases have shown good performance in completeness metrics, e.g., line/branch coverage and mutation testing; they are limited in correctness metrics, e.g., writing correct assertion statements.

% yibo: lack of the importance of CoTs before this paragraph to answer: Why complete focal methods and test cases **without CoTs** is not enough? Maybe can refer to some evalutaion results in our paper.
To overcome the challenge, we propose a new approach, named \approach{} to synthesize high-quality unit tests with corresponding CoTs based on self-debugging. Our key insight is that, although the correctness and quality of CoTs cannot be directly evaluated, they can be incrementally improved by including them in the self-debugging loop of test cases.
Based on the insight, we design our approach with two main techniques, guided test repair and CoT compression. 
First, guided test repair iteratively repairs generated test files through execution-driven self-debugging.
We execute a generated test in the automatically constructed environment. 
We select the most critical defect for self-debugging according to the execution results while generating a corresponding debugging CoT. 
Second, CoT compression consolidates the original generation CoT and the debugging CoT into a compact CoT that explains how the repaired test is derived from the focal file while removing transient debugging details. 
By formulating CoT construction as a rewriting process rather than direct regeneration, our approach preserves causal continuity across iterations and yields lightweight, high-quality training data for unit test generation.

% Based on the insight, our approach includes two main techniques, guided test repair and CoT compression. First, given a wrongly generated test case and the corresponding CoT (the original CoT) of the LLM, guided test repair guides the LLM to self-debug the test case and produce a debugging CoT that explains how to fix the test case. The debugging CoT can then correct and supplement the original CoT. Specifically, we conduct a heuristic self-debugging loop that includes error-self-debugging, failure-self-debugging, and coverage-self-debugging to get debugging CoTs that lead to test cases with high correctness and coverage. Second, CoT compression compresses the original CoTs and debugging CoTs into compressed CoTs that directly explain the generation of the correct test cases. The compressed CoTs are then good training data for UT generation tasks. 

To evaluate the effectiveness of our approach, we conduct a full process of training-data synthesis and supervised fine-tuning (SFT).
First, we collect 68,647 <focal method file, test file> pairs from an existing dataset, \textit{pymethod2test}~\cite{pymethods2test}. % yibo: why this dataset needs to be explained.
We then apply our self-debugging pipeline to these focal files to synthesize high-quality training data with CoTs for file-level UT generation, forming a dataset named \dataset{}. 
Second, we use these training data to perform SFT over Qwen2.5-Coder-32B-Instruct and DeepSeek-R1-Distilled-Qwen3-8B to validate the effectiveness of our synthesized data on a benchmark TestGenEval~\cite{testgeneval}. % yibo: any other benchmarks? does the capability of unit test generation can be translate to improvements in general programming tasks, e.g., code generation?
Our evaluation shows that the fine-tuned model achieves a pass rate of 36.17\% on test assertions, a branch coverage of 43.90\%, and a mutation score of 88.66\% while state-of-the-art commercial models such as o4-mini achieve only 27.23\%, 16.34\%, and 76.82\%, respectively. % yibo: Why GPT-o3-mini?

% yibo: 我建议给approach起名，而非给dataset起名，除非本文重点聚焦在dataset上面。
Our main contributions are as follows:
% \vspace{-0.1cm}
\begin{itemize}[leftmargin=2em]
    \item A new approach named \approach{} for distilling high-quality UT generation data with CoT.
    \item A new dataset named \dataset{} that includes 74,518 high-quality data for UT generation with CoT.
    \item A UT generation model distilled from DeepSeek-R1, achieving a pass rate of 36.17\% on test assertions, a branch coverage of 43.90\%, and a mutation score of 88.66\% whereas o4-mini achieves only 27.23\%, 16.34\%, and 76.82\%, respectively.  
\end{itemize}

%% file: background.tex
\section{Motivation}\label{sec: background}

In this section, we illustrate our motivation using a unit test (UT) generation task extracted from a real-world repository, \textit{nilearn}.\footnote{\url{https://github.com/nilearn/nilearn/tree/d3caac15d1e563c614c344e6fdfdcf246a1858af}}
\textit{nilearn} is a widely used repository with over one thousand GitHub stars and provides statistical and machine-learning tools.

\begin{figure}[t]
\begin{minipage}{0.42\textwidth}
    \centering
    % \label{listing:example-src-code}
\begin{lstlisting}[language=python, style=code, linebackgroundcolor={%
    \ifnum\value{lstnumber}>9\relax
      \ifnum\value{lstnumber}<16\relax
        \color{Yellow!90!white}%
      \fi
    \fi
    \ifnum\value{lstnumber}>18\relax
      \ifnum\value{lstnumber}<26\relax
        \color{YellowOrange!90!white}%
      \fi
    \fi
}]
import numpy as np
from scipy import ndimage, linalg
from .. import _utils
from .._utils.niimg import _get_data
...

def reorder_img(img, resample=None):
  affine = img.affine.copy()
  A, b = to_matrix_vector(affine)
  if not np.all((np.abs(A) > 0.001)
      .sum(axis=0) == 1):
    ... # The affine is not nearly diagonal
    return resample_img(img, 
      target_affine=target_affine,
      interpolation=resample)
  axis_numbers = np.argmax(np.abs(A), axis=0)
  data = _get_data(img)
  # swapping out-of-order axes
  while not np.all(np.sort(axis_numbers) 
    == axis_numbers):
    first_inversion = 
      np.argmax(np.diff(axis_numbers) < 0)
    axis1 = first_inversion + 1
    axis2 = first_inversion
    data = np.swapaxes(data, axis1, axis2)
    ... # Update affine
  return new_img_like(img, data, affine)
\end{lstlisting}
\subcaption{The source code under test (focal file)}
\end{minipage}
\hspace{0.5cm} % ← add horizontal space
\begin{minipage}{0.48\textwidth}
    \centering

\begin{lstlisting}[language=python, style=code, linebackgroundcolor={%
    \ifnum\value{lstnumber}>15\relax
      \ifnum\value{lstnumber}<28\relax
        \color{Yellow!90!white}%
      \fi
    \fi%
    \ifnum\value{lstnumber}>5\relax
      \ifnum\value{lstnumber}<15\relax
        \color{Green!70!white}%
      \fi
    \fi%
}]
import pytest
from numpy.testing import (assert_array_equal, ...)
from nilearn.image.resampling import (reorder_img, ...) 
...

def rotation(theta, phi):
    cos, sin = np.cos, np.sin
    a1 = np.array([[cos(theta), -sin(theta), 0],
                        [sin(theta), cos(theta), 0],
                        [0, 0, 1]])
    a2 = np.array([[1, 0, 0],
                        [0, cos(phi), -sin(phi)],
                        [0, sin(phi), cos(phi)]])
    return np.dot(a1, a2)
    
def test_reorder_img():
  data = rng.uniform(size=(5, 5, 5, 2, 2))
  affine = np.eye(4)
  affine[:3, -1] = 0.5 * np.array(shape[:3])
  ref_img = Nifti1Image(data, affine)
  for theta, phi in random.randint(4, size=(5, 2)):
    rot = rotation(theta*np.pi/2, phi*np.pi/2)
    b = 0.5 * np.array(shape[:3])
    new_affine = from_matrix_vector(rot, b)
    rot_img = resample_img(ref_img, ...)
    np.testing.assert_array_equal(rot_img.affine, ...)
    ... # more assertions
\end{lstlisting}
\subcaption{The test file from source code}
\end{minipage}
% \vspace{-0.2cm}
\caption{
The code snippet in \colorbox{Green!70!white}{Green} is helper function used in test cases. The code block in \colorbox{Yellow!70!white}{Light-Yellow} is a code branch and its corresponding test case.
The code snippet in \colorbox{YellowOrange!70!white}{Dark-Yellow} is an uncovered branch.
}
\label{fig: example code}
\end{figure}

\subsection{Task Formalization}
In this paper, we focus on file-level UT generation.
The input of the UT generation task is a source code file of focal methods (a.k.a., focal file in the rest of this paper) in a repository.
Figure~\ref{fig: example code} shows a snippet of the focal file (\CodeIn{nilearn/image/resampling.py}). 
This file contains utility methods for applying arbitrary affine transformations to images.
The output of the UT generation task is the corresponding test file for the focal file. 
Figure~\ref{fig: example code}(b) shows the test file (\CodeIn{nilearn/image/tests/test \_resampling.py}) for the given focal file in the original repository. 

A test file typically comprises three parts: (1) import statements (Lines 1-4), which import symbols from the focal file and other related files; (2) helper functions (highlighted in green), which provide supporting functionalities, e.g., multiplying two rotation matrices or creating an empty image for testing; and (3) a set of test methods (highlighted in light yellow), each of which evaluates a specific focal method through a sequence of operations and assertions.

% Unlike conventional method-level UT generation, we focus on file-level UT generation.
\textbf{Motivation for File-Level UT Generation.}
It requires a model to reason jointly about the focal file and its repository context, enabling it to construct correct import statements and implement necessary helper functions, rather than merely generating isolated test cases.
% A model must understand both the focal file and its repository context to construct correct import statements and implement necessary helper functions, rather than merely generating an isolated test case.
For example, the method \CodeIn{rotation} in Figure~\ref{fig: example code}(b) generates a three-dimensional rotation matrix and is reused across multiple test cases, including \CodeIn{test\_reorder\_img}.
In a method-level test generation setting, such a helper function is either assumed to be provided or inlined within individual test cases; however, both practices violate the test development setting or the requirement for modularization.

Additionally, existing LLMs obtain very high scores on method-level UT generation tasks but perform substantially worse at the file level.
For example, existing benchmarks~\cite{testgeneval, testeval} show that GPT-4o achieves 97.2\% branch coverage on method-level UT generation versus 35.2\% in a file-level setting. 
Therefore, our data-synthesis pipeline is designed specifically to improve the capabilities—such as constructing a comprehensive test file with correct imports and helper functions—where existing LLMs remain weak.

\subsection{Benefits of CoTs in UT Generation}
Note that existing work~\cite{yang2024chain, zhu2025uncertainty} demonstrates that chain-of-thoughts (CoTs) significantly improve LLM performance on code-related tasks. Listing~\ref{lst: r1 generate cot} illustrates how an existing LLM employs a CoT to produce unit tests. 
First (Lines 1–4), the CoT analyzes the focal file to identify the methods to be tested. 
Second (Lines 5–11), it inspects the repository and testing ecosystem, including testing frameworks (e.g., \textit{unittest} or \textit{pytest}) and relevant modules in \textit{nilearn}.
Third (Lines 12–14), it elaborates on the design of individual test cases for each focal method. 
For example, when generating tests for \CodeIn{reorder\_img}, the CoT considers two scenarios that correspond to the method’s usages.

When guided by an explicit and comprehensive CoT, LLMs are less likely to omit necessary cases and can produce more complete test files. 
Consequently, our objective is to construct a high-quality training dataset of <focal file, test file> pairs augmented with CoTs, thereby improving LLM performance on file-level UT generation.

\begin{figure}[t]
    \centering    
\begin{minipage}{1.0\textwidth}
    \centering
\begin{lstlisting}[label = lst: r1 generate cot, caption = The source code under test extracted from the repository anthem, style=cot]
We are writing unit tests for the module `nilearn.image.resampling`.
The goal is to test the functions: 
  - resample_img, resample_to_img, reorder_img, from_matrix_vector, 
  - coord_transform, get_bounds, get_mask_bounds, and the exception BoundingBoxError.
We will use:
  - unittest or pytest (but note: the test file should be self-contained and not require pytest if using
    unittest)
  - numpy for array operations
  - nibabel for creating Nifti images
  - and nilearn itself for some utilities.
  ... # Details about testing other methods
For `reorder_img`:
  1. Without resampling: we flip the x-axis and reorder without resampling, then check the affine has positive diagonal.
  2. With resampling: we apply a rotation and then reorder with resampling, then check the affine has positive diagonal.
\end{lstlisting}
\end{minipage}
\begin{minipage}{1.0\textwidth}
    \centering
\begin{lstlisting}[label = lst: qa gen cot, caption = The source code under test extracted from the repository anthem, style=cot]
We will structure our test file as follows:
  - Import necessary modules: os, copy, math, numpy, pytest, nibabel, and the functions from nilearn that we
    are testing.
  - Write helper functions if needed ((*@\colorbox{Red!90!white}{like the \textit{rotation} function in the ground truth}@*)).
  - Write individual test functions starting with "test_".
  However,(*@\colorbox{Red!90!white}{note that the ground truth test file is quite long}@*)and we are to write a unit test file that
    achieves high coverage.
...
\end{lstlisting}
\end{minipage}
% \vspace{-0.2cm}
% \caption*{
% say something
% }
\end{figure}

\textbf{Pitfalls of generating CoTs from ground-truth tests.} 
Given that most UT-generation datasets pair focal files with ground-truth tests, a natural idea is to prompt an LLM to generate a CoT from both the focal file and its corresponding test file. 
However, When exposed to the ground-truth test file, the model shifts from inferring tests to merely explaining an existing one, leading to omission of necessary intermediate reasoning steps and leakage of test-specific artifacts, which violates the real-world usage of a UT generation model.

Listing~\ref{lst: qa gen cot} illustrates the output DeepSeek-R1 when asked to directly generate reasoning for the <focal file, test file> pair in Figure~\ref{fig: example code}. 
Even when explicitly instructed not to reference the ground-truth test file, the CoT exhibits two clear limitations.
First, it incorporates elements that appear only in the test file.
For example, Line 4 mentions the helper function \CodeIn{rotation} without inferring logic from the focal file. 
Second, it outputs the final test-file structure directly, bypassing the intermediate analysis typically observed in valid CoTs (e.g., Lines 12–14 in Listing~\ref{lst: r1 generate cot}). 
These issues indicate that the model relies on the ground truth rather than performing genuine CoT reasoning over the focal file.

\subsection{Our Motivation}
Motivated by existing LLM technical reports~\cite{phi, phi15, seed2025seed}, we propose to synthesize both CoTs and test files, and simultaneously refine their quality through data-augmentation.
% rather than collecting test files directly from open-source repositories. 
% This approach addresses the main limitation: because the test file is not provided to the model during generation, the resulting CoTs reflect genuine reasoning grounded in the focal file rather than leaked explanations of ground-truth test files. 
% Moreover, subsequent data augmentation enables us to simultaneously refine the quality of both synthesized test files and their associated CoTs.
In the data-augmentation stage, we apply self-debugging~\cite{safe} to the synthesized test files and their associated CoTs.
For each generated test file, we assess its quality using both correctness and completeness metrics.
We then prompt the same LLM to conduct self-debugging on execution errors, assertion failures, insufficient coverage, and failure to kill mutants. 
After self-debugging, we compress the CoTs produced during both initial generation and the subsequent self-debugging process into a single, consolidated test-generation CoT.
This compression preserves consistency between the final CoT and the corrected test file while preventing uncontrolled context growth across multiple debugging turns~\cite{zhang2024imperative, laban2025llms}.
Thus, we maintain a bounded prompt/context size while preserving the causal chain of decisions needed for subsequent rounds of generation and debugging.

The key advantage of our self-debugging approach is that it converts the evaluation of CoT, which are inherently subjective and difficult to quantify, into the evaluation of executable test files, which can be measured objectively using clear correctness and completeness metrics.
An additional benefit is that the self-debugging pipeline might potentially improve test-file quality, particularly for tests that are initially low-quality or poorly maintained.

%% file: approach.tex
\section{Approach}~\label{sec: approach}
\begin{figure}
    \centering
    \includegraphics[width=\linewidth]{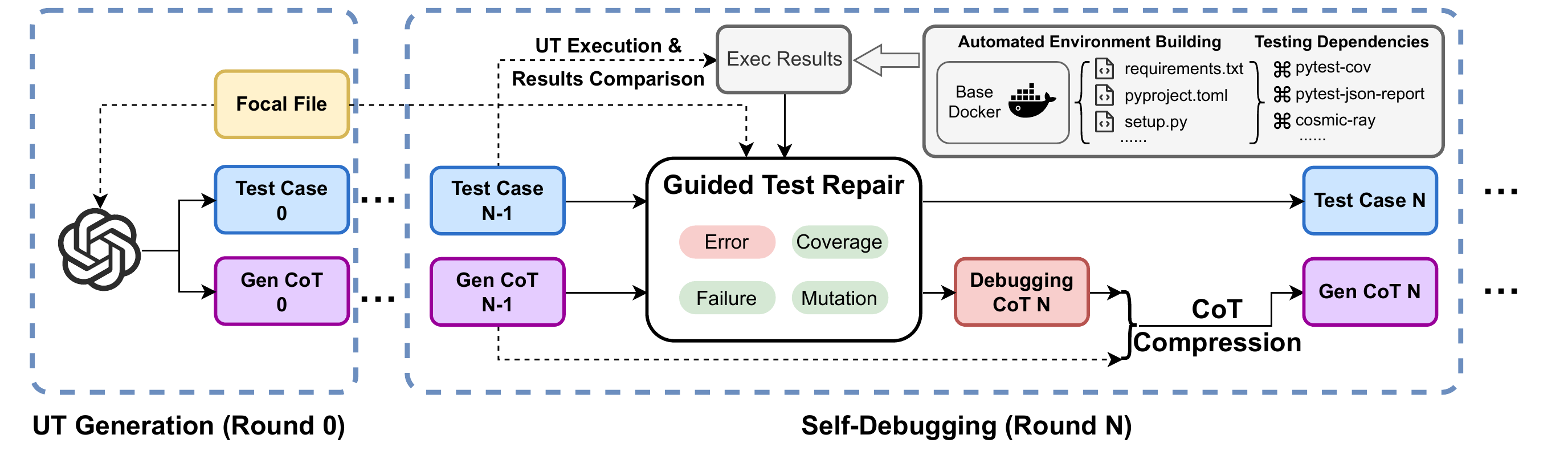}
    \caption{Overview of \approach{}}
    \label{fig: overview}
    % \vspace{-0.2cm}
\end{figure}

In this section, we present \approach{}, the first approach for synthesizing unit test (UT) generation data with high-quality CoTs via self-debugging.
The overall workflow of \approach{} is illustrated in Figure~\ref{fig: overview}. \approach{} consists of one round of UT generation followed by $n$ rounds of self-debugging, and each round of self-debugging involves two stages: guided test repair and CoT compression.
% \hzy{The self debugging in the figure should be put in the big rectangle. The original "LLM Self-debugging" should be changed to "guided test repair".}

In the first round of UT generation ($LLM UT Generation$), \approach{} takes a focal file from a code repository as input and outputs a corresponding test file $Test\textit{-}File_0$ along with its Chain-of-Thought (CoT) $Gen\textit{-}CoT_0$ that leads to the generated test file.

% \hzy{we need to show the following two stages in the original figure.}

Then in the following rounds, \approach{} performs self-debugging on the test files and the corresponding generated CoTs. Each round comprises two stages: the guided test repair stage and the CoT compression stage.
In the guided test repair stage, we prompt the same LLM to repair a specific correctness or completeness defect in the last round's generated test file, $Test\textit{-}File_{N-1}$, into an updated test file $Test\textit{-}File_{N}$. 
Concurrently, the LLM emits a debugging CoT $Debugging\textit{-}COT_{N}$ that documents the reasoning used to identify and fix the defect.

In the CoT compression stage, we merge the previous generation CoT, $Gen\textit{-}CoT_{N-1}$, with the newly generated debugging CoT, $Debugging\textit{-}CoT_{N}$, to a compact, unified $Gen\textit{-}CoT_{N}$. 
Intuitively, $Gen\textit{-}CoT_{N-1}$ describes how the LLM initially generated $Test\textit{-}File_{N-1}$ from the focal file, while $Debugging\textit{-}CoT_{N}$ describes how the LLM transformed $Test\textit{-}File_{N-1}$ into $Test\textit{-}File_{N}$ by addressing a specific defect. 
Their combination therefore captures the end-to-end reasoning from the focal file to $Test\textit{-}File_{N}$.

\subsection{Guided Test Repair}
In this stage, we execute the test file produced in the previous self-debugging round, denoted $Test\textit{-}File_{N-1}$, and use the observed execution results to guide repairs. 
This stage comprises two principal components: automated environment building and repair type selection.

\textbf{Automated environment building.}
To run generated tests and collect correctness and completeness results, we construct an isolated and reproducible execution environment for each repository. 
Given the large number of Python repositories (on the order of 500k)~\cite{tsakpinis2024analyzing}, we design an automated building strategy in this component with the following three steps. For each repository, first, we create a minimal Docker image that provides a stable runtime (Ubuntu 22.04 with Python 3.10). 
Second, we attempt to locate a standard build or dependency specification (for example, \CodeIn{requirements.txt} or \CodeIn{pyproject.toml}). 
If such a file is found, we install all dependencies using the repository’s specified toolchain.
Repositories without an executable build specification are excluded from further processing.
Third, we install additional testing and analysis tools required by our pipeline, e.g., \CodeIn{pytest-cov} for collecting coverage reports and \CodeIn{cosmic-ray} for mutation testing.

% Although prior work~\cite{swegym, swesmith} has constructed Docker images for a large number of Python repositories, we adopt our general on-demand building strategy primarily for efficiency.
% A Docker image for a Python repository can occupy 500~MB to 1~GB, making it prohibitively expensive to persist images for thousands of repositories. 
% Moreover, excluding a small fraction of repositories without executable build scripts is acceptable in a data synthesis task for LLM training, as most high-quality and widely used projects provide standard build configurations.

\textbf{Repair type selection.}
After executing $Test\textit{-}File_{N-1}$, we determine the most appropriate self-debugging action by analyzing the discovered execution defects. 
We classify execution defects into four main categories: (1) execution errors (e.g., import failures, runtime exceptions, syntax errors), (2) failed test assertions, (3) insufficient coverage, and (4) failure to kill mutants (low mutation score).
% Additionally, we treat execution errors as a high-priority type because successful execution is a prerequisite for assessing other correctness and completeness metrics.

Algorithm~\ref{alg: debugging selection} shows the pseudocode of our repair type selection algorithm. 
It first checks for execution errors (Lines 1–3) as successful execution is a prerequisite for assessing other correctness and completeness metrics. If such a defect exists, it selects this defect immediately. 
If not, the algorithm computes three evaluation metrics for the current test file: the assertion pass rate, the line coverage rate, and the mutation score, and converts them to a relative score to the ground-truth test (Lines 4-6). If all these scores are above 1, the self-debugging process stops, as the current test has reached the quality of the ground-truth test.

% them to the corresponding metric scores of the ground test \hzy{(Lines TODO-TODO, complete later)}. 
% Each metric is converted to a relative score that reflects the generated test’s performance compared to the ground-truth test file.

\begin{algorithm}[t]
	\SetKwData{Left}{left}\SetKwData{This}{this}\SetKwData{ast}{AST}\SetKwData{jdt}{jdt}
	\SetKwFunction{Union}{Union}\SetKwFunction{FindCompress}{FindCompress}
	\SetKwInOut{Input}{input}\SetKwInOut{Output}{output}
	
	\Input{$curTest$, a generated test file; $gtTest$: the ground-truth test file}
	\Output{the test file after self-debugging}
	\BlankLine
    \If{\mbox{\textbf{not}} Execution(curTest)}{
        $choice \gets exec$
    }
    \Else{
        $S_{pass} \gets Pass(curTest) / Pass(gtTest)$\;
        $S_{cov} \gets Coverage(curTest) / Coverage(gtTest)$\;
        $S_{mut} \gets Mutation(curTest) / Mutation(gtTest)$\;
        \If{$min(S_{pass}, S_{cov}, S_{mut}) \geq 1$}{
            \Return{\textbf{None}}
        }
        % \For{$idx \in [pass, cov, mut]$}{
        %     $s_{idx} \gets (S_{idx} + 1)^{-\alpha}$
        % }
        $choice \gets \mathop{argmin}\limits_{idx \in [pass, cov, mut]} \{S_{idx} \}$\;
    }
    $errorMsg \gets Construct(choice, curTest)$\;
    $newTest \gets Debug(curTest, ErrorMsg)$\;
    \Return{$newTest$};
\caption{Self-Debugging Selection}\label{alg: debugging selection}
\end{algorithm}

Otherwise, we then select the repair target by identifying the metric with the lowest relative score, since the lowest metric indicates the most critical defect in the generated test. 
The chosen metric determines the debugging type (e.g., repairing execution error/failure assertion, or improving coverage and mutation score).

After selecting the targeted prompt, we prompt the LLM to perform test repair on $Test\textit{-}File_{N-1}$, as illustrated in Listing~\ref{prompt: ut gen}. The LLM then generates a repaired test file $Test\textit{-}File_N$ along with the debugging CoT $Debugging\textit{-}CoT_N$.

\begin{figure}
    \centering   

\begin{minipage}{0.96\linewidth}
\begin{lstlisting}[caption = The Self-Debugging Prompt, label = prompt: ut gen, style=code, frame = shadowbox, linebackgroundcolor={%
    \ifnum\value{lstnumber}>2\relax
      \ifnum\value{lstnumber}<5\relax
        \color{Red!50!white}%
      \fi
    \fi%
    \ifnum\value{lstnumber}>5\relax
      \ifnum\value{lstnumber}<8\relax
        \color{Orange!50!white}%
      \fi
    \fi%
    \ifnum\value{lstnumber}>8\relax
      \ifnum\value{lstnumber}<12\relax
        \color{YellowOrange!50!white}%
      \fi
    \fi%
    \ifnum\value{lstnumber}>12\relax
      \ifnum\value{lstnumber}<15\relax
        \color{Yellow!50!white}%
      \fi
    \fi%
}]
The generated test Python file cannot be successfully executed.
(*@\color{gray!90!white} \# Prompt for execution error@*)
There exists errors in the test file, please help debug the test file according to the error messages.
The error messages are: {error_message}
(*@\color{gray!90!white} \# Prompt for assertion failure@*)
Some tests fail, please help debug the test file according to the failure messages.
The failure messages are: {failure_message}
(*@\color{gray!90!white} \# Prompt for insufficient coverage@*)
Some lines are not covered, please help improve the test file to increase code coverage.
Line {start_line} to {end_line} are not covered.
These lines are: ```python {missing_lines} ```
(*@\color{gray!90!white} \# Prompt for failure to kill mutants@*)
Some mutants are not killed in mutation testing, please improve the test file to increase mutation score.
The mutant that is not killed has the following diff: ```python {mutant_diff} ```
Improve the thinking process so the unit tests are more complete and robust and output the unit test Python
file in this format: ```python Unit test Python code (file level) ```
\end{lstlisting}

% \vspace{-0.2cm}
% \caption*{
% say something
% }
\end{minipage}
\end{figure}

% \begin{figure}
%     \centering    
% \begin{lstlisting}[caption = The Self-Debugging Prompt, label = prompt: ut gen, style=cot]
% Below is a code file:
% ```python
% {focal_file_content}
% ```
% The code file is called: {focal_file_path}
% Your job is to output a corresponding unit test file that obtains high coverage and invokes the code under test.
% Here are some examples of how to import {focal_file_path}, (you can use these as reference)
% ```python
% {imports}
% ```
% Each unit test must be a function starting with test. 
% Include all your test imports and setup before your first test. 

% Do not run the tests in the file, just output a series of tests.
% Do not include a main method to run the tests.
% Only output the unit test Python file in this format:
% ```python
% Unit test Python code (file level)
% ```
% \end{lstlisting}

% % \vspace{-0.2cm}
% % \caption*{
% % say something
% % }
% \end{figure}

% First, given a wrongly generated test case and the corresponding CoT (the original CoT) of the LLM, CoT-involved self-debugging guides the LLM to self-debug the test case and produce a debugging CoT that explains how to fix the test case. The debugging CoT can then correct and supplement the original CoT. Specifically, we conduct a heuristic self-debugging loop that includes error-self-debugging, failure-self-debugging, and coverage-self-debugging to get debugging CoTs that lead to test cases with high correctness and coverage.

\subsection{CoT Compression}
The CoT compression stage produces a compact CoT $Gen\textit{-}CoT_{N}$ that concisely explains how the LLM generates the repaired test file $Test\textit{-}File_{N}$ from the focal file.
The compressed CoT both (a) preserves the essential reasoning steps required to reproduce the generation and (b) removes debugging details ($Debugging\textit{-}CoT_N$) introduced in the guided test repair stage, thereby yielding lightweight and high-quality training data for UT-generation.

We model CoT compression as a rewriting task where the LLM is instructed to integrate the modifications in $Debugging\textit{-}CoT_{N}$ into the original generation CoT $Gen\textit{-}CoT_{N-1}$ with a specialized prompt (Listing~\ref{prompt: cot compression}).
The prompt provides the two CoTs as inputs and includes explicit guidelines instructing the LLM to avoid mentioning any intermediate notations introduced by the prompt, ensuring that the output is a clean and self-contained reasoning trace.

\textbf{Difference from direct CoT generation based on test files.}
A natural question is why we do not directly prompt the LLM to generate a new CoT from the focal file and test file. 
The main reason is that our design reformulates CoT construction from a drifting generation task into a consistent rewriting task. 
The LLM is tasked solely with updating this trace according to the newly identified debugging steps rather than generating it from scratch. 
This design helps preserve causal continuity across iterations and thus avoids unnecessary variance in newly generated CoTs.
% Directly generating CoTs based on test files requires LLMs to synthesize a reasoning trace from scratch, which is prone to drift and inconsistency.
% Since our CoT compression stage already starts from a genuine CoT that correctly explains the prior test generation, the LLM is tasked solely with updating this trace according to the newly identified debugging steps. 
% This design helps preserve causal continuity across iterations and thus avoids unnecessary variance in newly generated CoTs.

\begin{figure}
    \centering  
\begin{minipage}{0.95\linewidth}
\begin{lstlisting}[caption = The CoT Compression Prompt, label = prompt: cot compression, style=cot]
You are a senior test engineer. You will be given:
- R0: original reasoning that produced a first version of a test file.
- R1: debugging reasoning that produced an improved version of the test.
- T1: the improved test file.

GOAL:
Merge R0 and R1 into a comprehensive reasoning step R2 that explains T1.

Guidelines:
- Do NOT reference or mention R0, R1, and T1 anywhere.
- Your output reasoning step should be the same as generating T1 based on a single focal file.
- Your output should focus on the thinking process that led to T1.
- Mainly following the structure of R0, but also incorporating relevant insights from R1.

The original reasoning R0 is: ``` {r0_thinking} ```
The debugging reasoning R1 is: ``` {r1_thinking} ```
The improved test file T1 is: ```python {test_file_content} ```
\end{lstlisting}
% \vspace{-0.2cm}
% \caption*{
% say something
% }
\end{minipage}
\end{figure}

%% file: evaluation.tex
\section{Evaluation}~\label{sec: evaluation}
Our evaluation answers the following four Research Questions (RQs):
\begin{enumerate}
    \item What is the quality of our synthesized test files?
    \item What are the benefits of each round's self-debugging on our synthesized test files?
    \item To what extent do our synthesized test files improve LLM's UT generation capabilities?
    \item To what extent do our synthesized CoTs improve LLM’s UT generation capabilities?
    % \item What are the performance differences of our fine-tuned models across different repositories or functionalities?
\end{enumerate}

\subsection{Evaluation Setup}
Our evaluation is comprised of the data synthesis stage and the model evaluation stage. 
In the data synthesis stage, we use a source dataset of test-file generation tasks and enhance it to a high-quality synthesized dataset with CoTs using our data synthesis approach.
In the model evaluation stage, we fine-tune an LLM based on our synthesized dataset and evaluate the model's UT generation capabilities on benchmarks.

% \hzy{I think we should use Repo-Smith as the approach for generating a dataset with CoT and use names like pymethod2testCoT or sth else as the generated high-quality dataset.}

% \quad 
\subsubsection{Dataset}
We choose pymethod2test~\cite{pymethods2test} as our source dataset for the data synthesis stage, and TestGenEval~\cite{testgeneval} as our evaluation benchmark for the model evaluation stage.
\begin{itemize}
    \item \textbf{pymethod2test}~\footnote{\url{https://zenodo.org/records/14264519}} collects 1,289,630 test files from 88,846 open-source Python repositories. It uses a series of manually-designed heuristics to map focal files and test files. Such data scale is sufficient for data synthesis.
    \item \textbf{TestGenEval}~\footnote{\url{https://huggingface.co/datasets/kjain14/testgeneval}} is a UT generation benchmark extracted based on SWEBench~\cite{swebench}. TestGenEval comprises 68,647 tests from 1,210 code and test file pairs across 11 well-maintained Python repositories (in SWEBench). Compared with other UT generation benchmarks~\cite{testeval}, which focus on generating test cases for a focal method, TestGenEval mainly focuses on file-level test generation. 
    % Such a design comes from its evaluation results that existing LLMs, e.g., GPT-4o and LLaMa3.1-8B, have already achieved a high performance in method-level test generation.
\end{itemize}

\subsubsection{Metrics}~\label{sec: metrics}
We evaluate the quality of generated tests (including the test files in the synthesized dataset and the test files generated by the fine-tuned models) on both \textit{correctness} and \textit{completeness} perspectives~\cite{testgeneval, safe, lahiri2024evaluating}.

\textit{Correctness} means that a correct focal file should pass all test cases. 
We evaluate the correctness of a given test file based on its execution rate and pass rate.
The execution rate is defined as whether a given test file is successfully executed, e.g., without syntax/import errors.
The pass rate is defined as the portion of passed test cases when testing the focal file. If a test file cannot be executed, its pass rate is defined as 0.
Specifically, we use pytest~\footnote{\url{https://docs.pytest.org/en/stable/}}, the most widely-used Python testing framework, to collect test execution results.

\textit{Completeness} means that any incorrect focal file should not pass all test cases. 
We evaluate the completeness of a test file via line/branch coverage~\cite{whalen2006coverage, aghababaeyan2023black} and mutation testing~\cite{baker2012empirical, kaufman2022prioritizing}.
Line/branch coverage measures the portion of a test file's executed code lines/branches, which is widely used as a necessary condition of test completeness.
Mutation testing measures how much proportion of mutant focal files (with manually-injected faults) can be detected by test cases, thus indicating the capability of test cases in detecting incorrect focal files in the real world.
To keep aligned with existing work~\cite{testgeneval}, we use cosmic-ray~\footnote{\url{https://cosmic-ray.readthedocs.io/en/latest/}} for mutation testing.

\subsubsection{Evaluated LLMs}
In the data synthesis stage, we use DeepSeek-R1~\cite{deepseekr1} for data synthesis.
% To validate the effectiveness of our synthesized data, 
In the model evaluation stage, we fine-tune and evaluate the effectiveness of Qwen2.5-Coder-32B-Instruct and DeepSeek-R1-Distilled-Qwen3-8B.
Additionally, we take DeepSeek-R1, o1/o3/o4-mini (three representative reasoning models proposed by OpenAI) as evaluation baselines.

\begin{itemize}
    \item \textbf{DeepSeek-R1~\cite{deepseekr1}}. DeepSeek-R1 is one of the state-of-the-art (SOTA) open-source reasoning models. It has 671B parameters and active 37B of them. It has been widely used in synthesizing high-quality training data~\cite{seed2025seed, opencodereasoning, opencodereasoning2}.
    \item \textbf{o1-mini/o3-mini/o4-mini~\cite{o4-mini}}. They are compact versions of OpenAI's o1/o3/o4 models, two SOTA reasoning models released by OpenAI. They are more intelligent than the previously widely evaluated GPT-4o and is designed to achieve higher reasoning performance utilizing fewer computational resources.
    % \item \textbf{GPT-o1-mini~\cite{TODO}}. 
    \item \textbf{Qwen2.5-Coder-32B-Instruct~\cite{qwen25coder}}. It is one of the widely-used Code LLMs. It uses 5.2 Trillion tokens during pre-training, including 70\% code, 20\% general text, and 10\% math corpus.
    \item \textbf{DeepSeek-R1-Distilled-Qwen3-8B}. It distills the enhanced CoTs from the upgraded DeepSeek-R1-0528. It is now stronger in complex reasoning, less hallucination, and better function calling compared to original Qwen3-8B. The resulting model achieves SOTA performance on math and code benchmarks, outperforming reasoning models with similar or even larger parameter scales.
\end{itemize}

\subsubsection{Training and Inference Hyper-Parameters.}
We evaluate the effectiveness of \dataset{} by fine-tuning the Qwen2.5-Coder-Instruct model with 32B parameters and DeepSeek-R1-Distilled-Qwen3 with 8B parameters.
These models are widely used to evaluate the effectiveness of synthesized long CoT data~\cite{opencodereasoning, opencodereasoning2}.

The models were trained for 2 epochs on NVIDIA H200-141GB
 GPUs, using the AdamW optimizer with a batch size of 128 and
 a maximum sequence length of 32,768.
We set the learning rate as 2e-5 and employed a Cosine Annealing scheduler with a warmup ratio of 0.03, 
We use the final checkpoint for evaluation.

Additionally, we adopt greedy decoding (i.e., use LLMs to generate only one answer on all inputs) during evaluation on fine-tuned models. 
The rationale behind this step is to reflect the real-world usage of a UT generation model, e.g., the GitHub Copilot extension outputs only one test case in a user's developing environment.

\subsection{RQ1: What is the quality of our synthesized test files?}
In this research question, we apply our self-debugging framework to \textit{pymethod2test} to synthesize unit-test generation data with extended Chains-of-Thoughts (CoTs).
We use the evaluation metrics defined in Section~\ref{sec: metrics} to assess the quality of the synthesized test files. 
We compare metric results across three categories of test files: those collected from source code, those directly generated by DeepSeek-R1, and those produced after five rounds of self-debugging. Additionally, we report the metric results for our synthesized test files under various ablation settings, including self-debugging without mutation debugging, coverage debugging, failure debugging, and their combinations.

\textbf{Main results.} 
Table~\ref{tab: main debugging strategy} summarizes the metric results of our synthesized test files on \textit{pymethod2test}. 
Compared with both the source-code tests in \textit{pymethod2test} and the test files directly generated by DeepSeek-R1, the test files produced by our self-debugging pipeline achieve substantially higher scores on nearly all metrics. Specifically, our approach attains an execution rate of 73.17 and a pass rate of 56.60, whereas the best corresponding baseline scores are only 46.21 and 26.82, respectively.

\textbf{Comparison between source code and directly-generated test files.}
An important observation is that the source-code tests in \textit{pymethod2test} and the test files directly generated by DeepSeek-R1 (Generate R0) yield similar metric scores. 
This similarity mainly stems from the limited quality of existing test-generation datasets, such as \textit{method2test} and \textit{pymethod2test}. 
Prior work~\cite{cleantest} highlights this issue and applies syntax-based filters to remove low-quality test cases, which inevitably introduces a trade-off in dataset size.
In contrast, our \approach{} method employs self-debugging to synthesize high-quality test data while preserving dataset scale, thereby offering greater potential benefits for model training.

\textbf{Contribution of Various Debugging Strategies.}
Table~\ref{tab: main debugging strategy} also reports the metric scores under different self-debugging strategies.
(1) Compared with the setting without mutation debugging, we observe almost the same scores on all metrics, suggesting that mutation debugging contributes relatively little to either correctness or completeness.
(2) Relative to the setting without coverage debugging, line and branch coverage show small decreases, whereas the execution rate and pass rate remain unchanged, suggesting that coverage debugging improves completeness while preserves correctness.
(3) When compared with the setting without mutation, coverage, and failure debugging simultaneously, all metrics exhibit substantial declines, and only the execution rate remains similar, which is mainly credited to error debugging.
Overall, these results indicate that failure debugging contributes most significantly to improving the quality of synthesized test cases, followed by coverage debugging, with mutation debugging contributing the least.

% \textbf{Whether the test files can be induced by CoTs?}
\textbf{Improvement on CoTs after Self-debugging.}
Given that our test cases demonstrate improvements in both correctness and completeness metrics, we further examine the impact of CoTs.
Table~\ref{tab: cot induced dataset} compares the quality of directly generated test files (Round 0) with test files generated using Round-5 CoTs as prefixes. 
Moreover, the self-debugging results reported in Table~\ref{tab: main debugging strategy} are substantially higher than any configuration that directly uses LLMs for generation.
When provided with Round-5 CoTs, we observe a slight decrease in execution and pass rates, accompanied by an increase in line and branch coverage. 
This difference arises from a trade-off between correctness and completeness. 
Specifically, CoTs refined through self-debugging are more comprehensive---they plan additional test cases to improve coverage, causing DeepSeek-R1 to produce more errors and failures.
These findings indicate that, without the self-debugging pipeline, the LLM alone cannot generate test files of comparable quality along with their corresponding CoTs.

\begin{tcolorbox}[colback=lightgray!25!white, colframe=gray, title=\textbf{Summary of RQ1}]
\textbf{Self-debugging's effectiveness:} 
Our self-debugging pipeline improves the quality of test files in both correctness and completeness metrics. 
Without the self-debugging pipeline, the LLM alone cannot generate test files of comparable quality.

\textbf{Contributions of debugging strategies:}
Failure debugging provides the largest improvement, followed by coverage debugging, with mutation debugging contributing the least.

\end{tcolorbox}

\begin{table}[t]
\small
\centering
\caption{The Metric Results of Synthesized Test Files (on pymethod2test) under Various Debugging Strategies}
\label{tab: main debugging strategy}
\begin{threeparttable}
\begin{tabular}{lrcccccccr}
\toprule
& \multicolumn{4}{c}{Number of Test Cases} & \multicolumn{5}{c}{Evaluation Metrics} \\
\cmidrule(lr){2-5}\cmidrule(lr){6-10}
                    & \#tests & \#passed & \#failed & \#error & exec & pass & $Cov_{l}$ & $Cov_{b}$ & $S_{mut}$ \\
\midrule
Source-Code        & 8.31  & 6.87   & 0.91   & 0.39  & 46.03        & 36.82      & 49.23          & 43.33            & 86.70          \\
\midrule
Generate (R0)      & 6.69  & 4.48   & 2.14   & 0.16  & 46.21        & 29.65      & 49.54          & 33.90            & \textbf{98.43}           \\
Self-Debugging (R5)       & 12.60 & \textbf{9.96}   & 2.09   & 0.42  & \textbf{73.17}        & \textbf{56.60}      & \textbf{66.75}          & \textbf{52.90}            & 91.90           \\
\quad - w/o Mut             & 12.58 & 9.93   & 2.09   & 0.42  & \textbf{73.17}        & \textbf{56.60}      & \textbf{66.75}          & \textbf{52.90}            & 91.69           \\
\quad - w/o Cov             & 12.38 & 9.68   & 2.15   & 0.42  & \textbf{73.17}        & 56.07      & 66.05          & 51.56            & 91.86           \\
\quad - w/o Mut/Cov         & 12.36 & 9.65   & 2.15   & 0.42  & \textbf{73.17}        & 56.07      & 66.04          & 51.56            & 91.63           \\
\quad - w/o Mut/Cov/Failure & 11.02 & 6.61   & 3.85   & 0.43  & 73.16        & 42.35      & 58.94          & 41.69            & 96.52           \\
\bottomrule
\end{tabular}
\begin{tablenotes}
    \item $Cov_{l}$ and $Cov_{b}$ are line coverage and branch coverage, respectively.
\end{tablenotes}
\end{threeparttable}
\end{table}

\begin{table}[t]
\small
\centering
\caption{The Execution Results and Evaluation Metrics of Test Files under Various Debugging Strategies}
\label{tab: cot induced dataset}
\begin{threeparttable}
\begin{tabular}{lrcccccccr}
\toprule
                    & \#tests & \#passed & \#failed & \#error & exec & pass & $Cov_{l}$ & $Cov_{b}$ & $S_{mut}$ \\
\midrule
Source-Code        & 8.31  & 6.87   & 0.91   & 0.39  & 46.03        & 36.82      & 49.23          & 43.33            & 86.70          \\
\midrule
Generate (R0)       & 6.69  & 4.48   & 2.14   & 0.16  & 46.21        & 29.65      & 49.54          & 33.90            & \textbf{98.43}           \\
Generate (R5's CoT)       & 7.71 & 4.89   & 2.85   & 0.12  & 41.27        & 23.63      &  52.04         & 35.96            & 91.90           \\
\bottomrule
\end{tabular}
\begin{tablenotes}
    \item Generate (R5's CoT) indicates that using Round 5's Chain-of-Thought as generation prefix to generate only test files, i.e., focal file + CoT $\rightarrow$ test file.
\end{tablenotes}
\end{threeparttable}
\end{table}

\subsection{RQ2: What are the benefits of each self-debugging round on the quality of the synthesized test files?}

In this research question, we evaluate the benefits of our core component, the self-debugging module, from two perspectives: (1) the acceptance rate (i.e., the proportion of test files included in the final synthesized dataset) of each round’s output test files, and (2) the quality of the test files generated in each round.

\begin{figure}[t]
\centering
\includegraphics[width=1\linewidth]{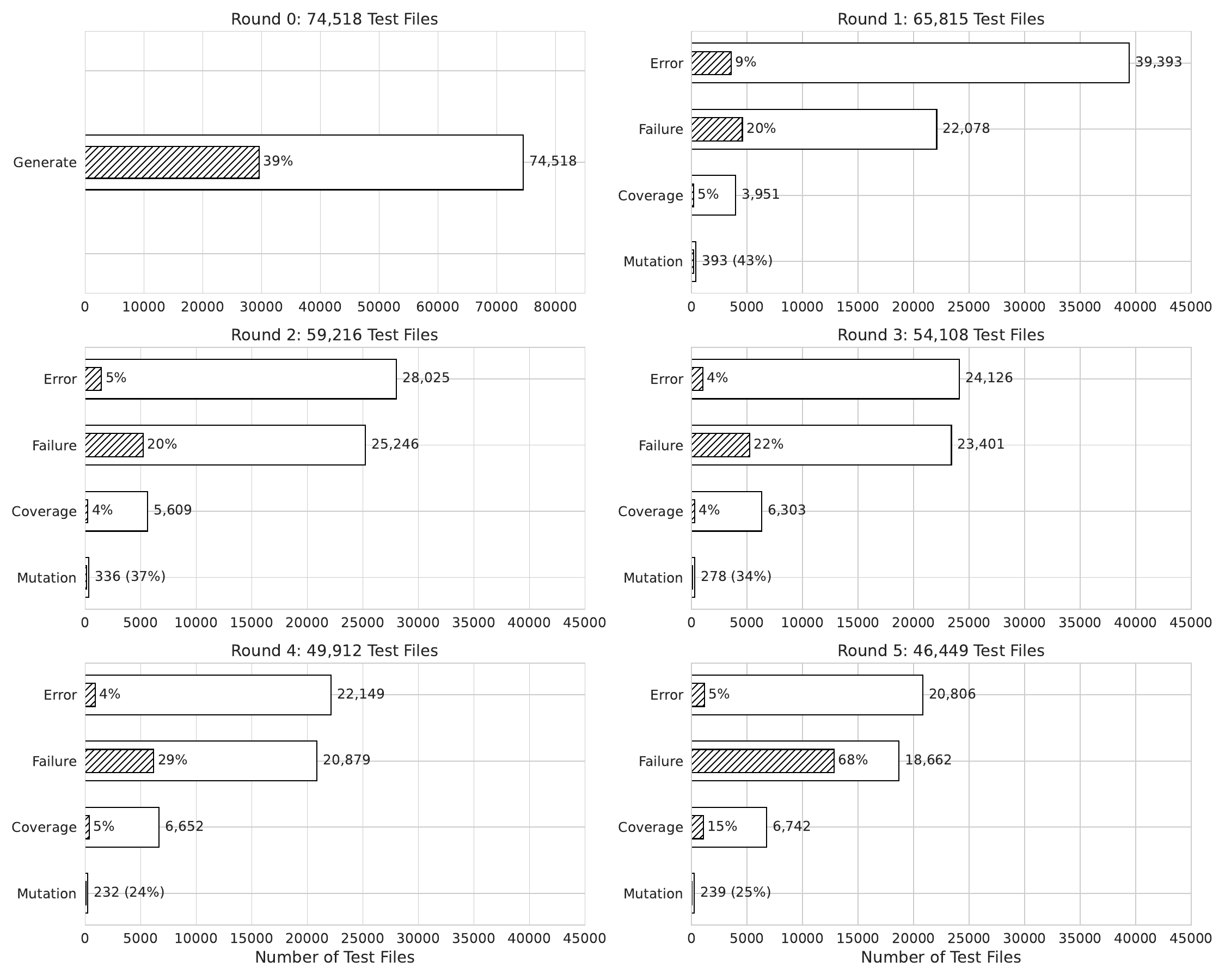}
\caption{The Distribution of Each Round's Debugging Types and Accept Rates}
\label{fig: self-debugging number}
% \vspace{-0.4cm}
\end{figure}

To assess acceptance rates, we report the number of debugging attempts issued in each round and the proportion of accepted test files. This acceptance ratio reflects each round's contribution to the final synthesized dataset.
To evaluate the test file's quality, we also apply the metrics defined in Section~\ref{sec: metrics} under two conditions: best of the first K rounds and only the K-th round. 
For a given focal file, best of the first K rounds denotes selecting the highest-quality test file generated within the first K self-debugging iterations, whereas only the K-th round refers to the test file generated specifically in round K. 
By comparing these two settings, we examine whether test-file quality tends to degrade over extended self-debugging and, consequently, identify when it may be beneficial to terminate further self-debugging rounds.

\textbf{Acceptance rate of each round’s test files.}
Figure~\ref{fig: self-debugging number} presents the total number and acceptance rate of test files generated in each self-debugging round, revealing three key findings.
(1) Failure debugging consistently achieves the highest acceptance rate (at least 20\%) across all rounds, whereas the rest of the debugging types remain below 10\%, indicating that failure debugging contributes the most to the end-to-end synthesized dataset.
(2) The volume of error debugging decreases sharply after the first round, while the volume of failure debugging remains relatively stable.
This trend suggests that a SOTA LLM (DeepSeek-R1) is effective at resolving execution errors but remains limited in generating correct test cases, particularly those requiring accurate test assertions.
(3) Approximately 11.7\% of directly-generated (Round 0) test files (decreasing from 74,518 of Round 0 to 65,815 of Round 1) are already better than ground-truths while 39\% of directly generated test files cannot be improved after self-debugging.

% cannot be improved after round 0 (direct generation). 
% \hzy{I didn't understand this, need to fix}
% \tianyu{directly generated test files are highly effective, 11\% are already better than ground-truths,
% 39\% cannot be improved after self-debugging.
% }
% The primary reason is that 66.5\% of these test files remain non-executable even after five rounds of error debugging, resulting in their classification as accepted in round 0.
% When excluding these inherently non-executable cases, our self-debugging pipeline yields substantial improvements and contributes to the quality of the synthesized dataset.

\begin{figure}[t]
\begin{minipage}{0.48\textwidth}
    \centering
    \includegraphics[width=1.\linewidth]{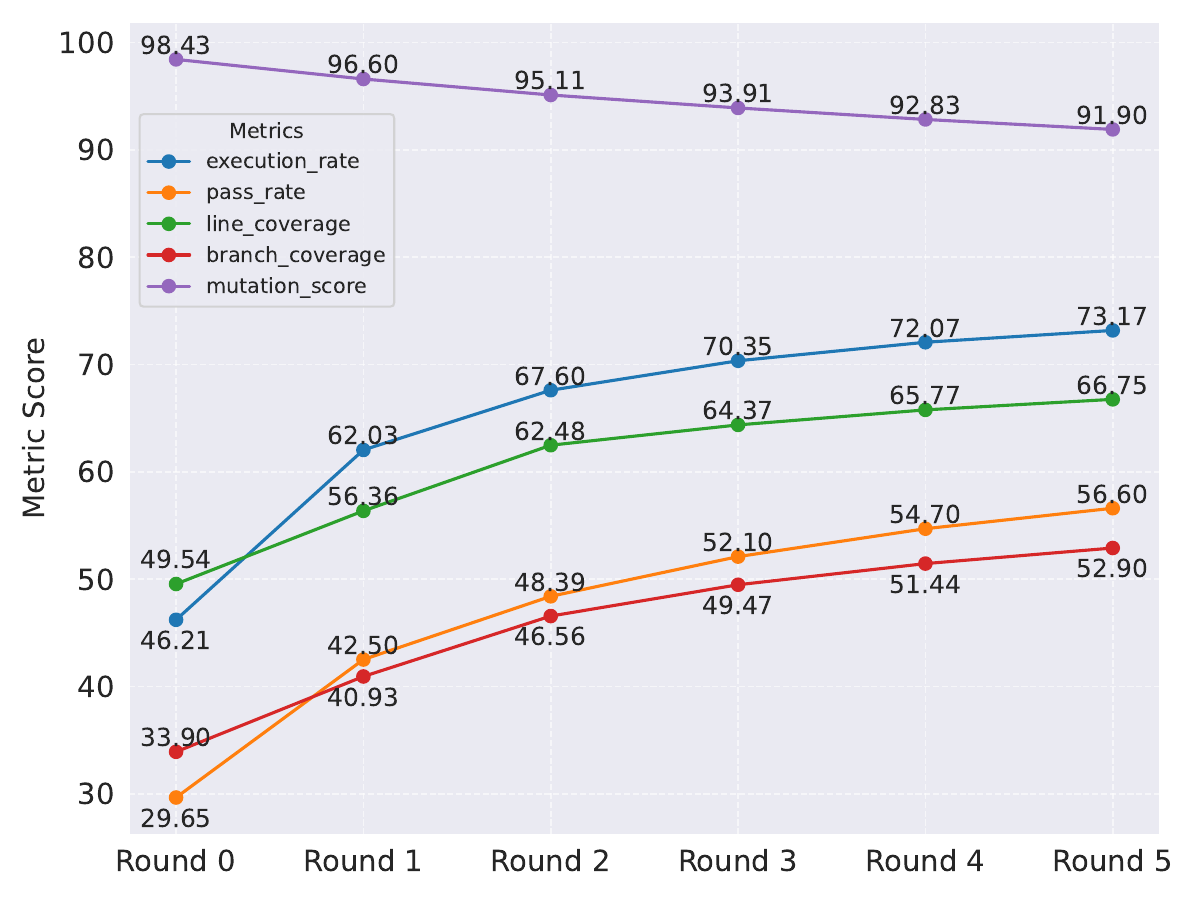}
    \captionsetup{justification = centerlast}
    \subcaption{Best of the first K Rounds}
    \label{fig: pymethod score first k}
\end{minipage}
\begin{minipage}{0.48\textwidth}
    \centering
    \includegraphics[width=1\linewidth]{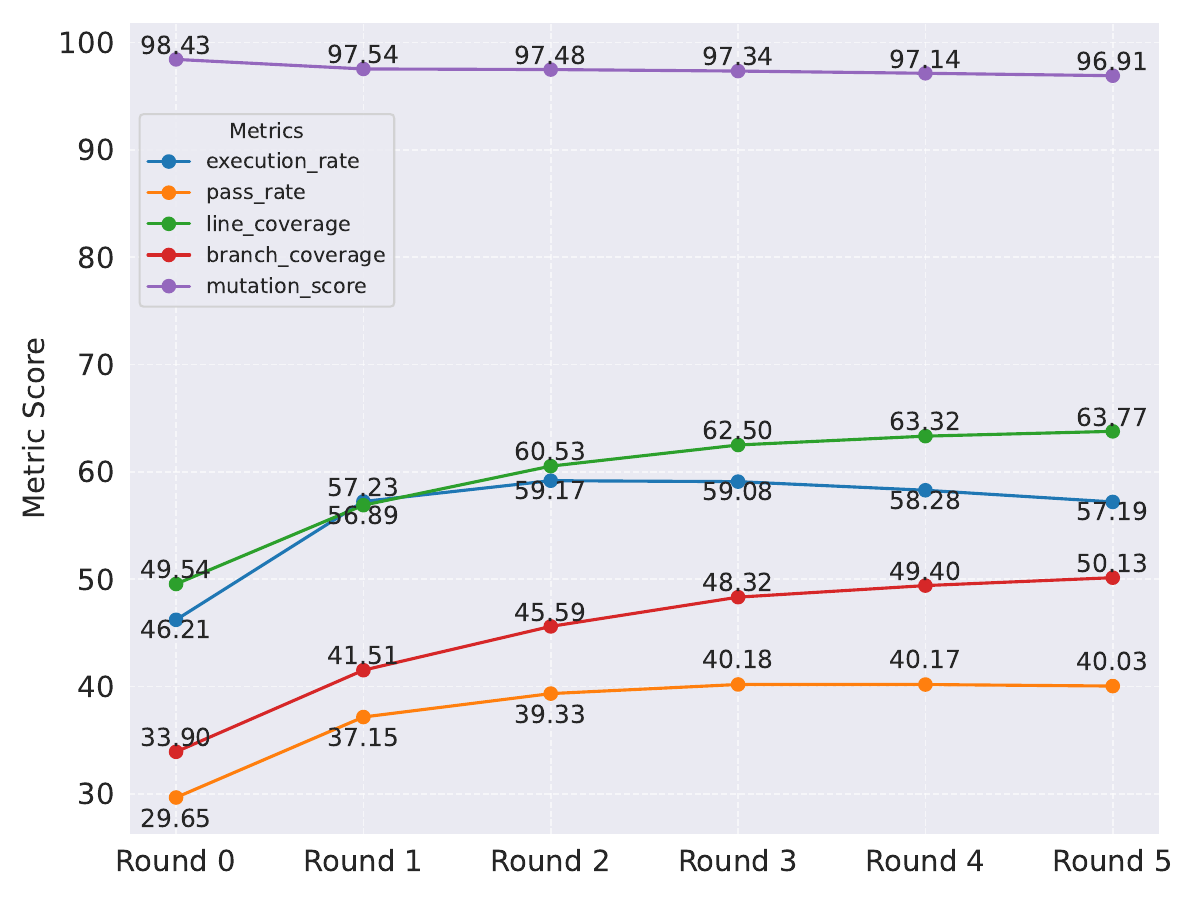}
    \captionsetup{justification = centerlast}
    \subcaption{Only the K-th Round}
    \label{fig: pymethod score only k}
\end{minipage}
% \label{fig: pymethod score}
\caption{The Metric Scores During Each Round's Self-Debugging}
% \vspace{-0.2cm}
\end{figure}

\textbf{Test-file quality over self-debugging rounds.}
Figure~\ref{fig: pymethod score first k} illustrates how our metric scores evolve across one round of generation and five rounds of self-debugging.
All evaluation metrics consistently increase as the number of debugging rounds grows. For instance, the execution rate improves by 15.82 (from 46.21 to 62.03), and the pass rate increases by 12.85 (from 29.65 to 42.50).
However, the increase from Round 2 to Round 5 is substantially smaller than in earlier rounds (0 to 2). 
This trend supports our decision to stop self-debugging after five rounds. 
% If computational or time resources are limited, using fewer self-debugging rounds remains a practical and effective option.

% \hzy{I think the following conclusion is straight-forward, maybe we can remove it? Maybe figure4a is already sufficient to show that the later rounds cannot improve much, where we can explain that it's the matter of the hard-to-solve cases.}
% \textbf{Drop in correctness scores after three self-debugging rounds.}
Figure~\ref{fig: pymethod score only k} presents the quality of the test files produced in the setting of only the K-th round.
A notable observation is that the correctness metrics (both execution rate and pass rate) begin to decline after the third self-debugging round. The primary reason is that the ``easy'' cases are largely resolved in earlier rounds, leaving only the more ``difficult'' cases in later rounds. 
As shown in Figure~\ref{fig: self-debugging number}, the number of remaining test files decreases from 65,815 in the first round to 46,449 in the fifth round, reflecting the increasing concentration of harder-to-debug cases.

\begin{tcolorbox}[colback=lightgray!25!white, colframe=gray, title=\textbf{Summary of RQ2}]
\textbf{Self-debugging is effective, with failure debugging contributing the most.}
Across all rounds, the self-bugging pipeline increases the quality and accepted number of test files, while failure debugging consistently shows the highest acceptance rate and contribution to the final synthesized dataset.

\textbf{Substantial improvements in early rounds with diminishing gains in later rounds:}
The first two self-debugging rounds account for most quality improvements and accepted test files, while gains beyond round 3 become marginal and correctness metrics decline, as remaining cases are increasingly difficult to debug.

% \textbf{Substantial improvements in early rounds:}
% All evaluation metrics rise notably from the initial generation to the first two self-debugging rounds, indicating that early-stage debugging effectively corrects execution issues and enhances overall test-file quality.

% \textbf{Improvements diminish in later rounds due to increasingly difficult cases:}
% After round 3, correctness metrics begin to decline because easier cases have already been resolved, leaving only harder-to-debug test files; thus, later rounds yield limited additional benefit.
\end{tcolorbox}

\subsection{RQ3: To what extent do our synthesized test files improve LLM's UT generation capabilities?}~\label{sec: eval fine tune}

This research question evaluates the end-to-end effectiveness of our synthesized UT-generation data.
We fine-tune two widely used open-source LLMs: Qwen2.5-Coder-32B-Instruct~\cite{qwen25coder} and DeepSeek-R1-Distill-Qwen3-8B~\cite{deepseekr1}, and assess all fine-tuned models on TestGenEval~\cite{testgeneval}.
To further investigate the impact of our self-debugging pipeline, we examine three types of SFT data: (1) the original source code from pymethod2test, (2) test files directly generated by DeepSeek-R1 (Round-0 data), and (3) test files produced after self-debugging (Round-5 data).

\textbf{Main results on TestGenEval.}
Table~\ref{tab: testgeneval main} presents the metric results of our fine-tuned models on TestGenEval.
Compared with the base LLMs, models fine-tuned on self-debugging data exhibit substantial improvements across all metrics.
Among all correctness and completeness metrics, Qwen2.5-Coder-32B-Instruct achieves an average improvement of 71\%.
Moreover, models fine-tuned on self-debugging data consistently achieve the highest scores among the three SFT settings.
For instance, when fine-tuned from Qwen2.5-Coder-32B-Instruct, it attains an execution rate of 68.13 and a pass rate of 36.17. In contrast, one of the best commercial models, o3-mini, attains only 52.50 and 36.72 on these metrics, respectively.

\begin{table}[t]
\small
\centering
\caption{The Execution Results and Evaluation Metrics on TestGenEval}
\label{tab: testgeneval main}
\begin{tabular}{lrrccccccr}
\toprule
                    & \#tests & \#passed & \#failed & \#error & exec & pass & $Cov_{l}$ & $Cov_{b}$ & $S_{mut}$ \\
\midrule
\textit{Results of Baseline Models:}\\
Ground-Truth              & 61.36 & 60.12  & 0.00   & 0.00  & 98.75        & 96.72      & 66.78    & 57.32      & 70.96          \\
Deep-Seek-R1              & 17.73 & 11.86  & 3.61   & 0.28  & 78.13        & 41.79      & 45.39    & 41.95      & 98.53           \\
o1-mini  & 9.58 & 8.11   & 1.03   & 0.04  & 46.88        & 26.30      & 41.79    & 28.54      & 78.01          \\
o3-mini               & 11.28 & 9.85   & 1.04   & 0.01  & 52.50        & 36.72      & 42.37    & 30.73      & 70.43          \\
o4-mini               & 10.20 & 8.90   & 1.00   & 0.00  & 46.25        & 27.23      & 40.99    & 16.34      & 76.82          \\
\midrule
\multicolumn{4}{l}{\textit{Results on Qwen2.5-Coder-32B-Instruct:}} \\
Raw Model   & 14.68 & 9.35   & 2.60   & 0.23  & 56.25        & 25.01      & 31.03    & 13.17      & 85.02          \\
 + SFT (Source Code)      & 5.65  & 3.59   & 1.19   & 0.03  & 34.38        & 9.12       & 25.54    & 19.34      & \textbf{98.79}           \\
 + SFT (R0 Data)     & 16.37 & 12.32  & 1.95   & 0.45  & 58.13        & 31.63      & 33.13    & 40.49      & 88.29          \\
 + SFT (R5 Data) & 18.78 & 12.93  & 3.66   & 0.42  & \textbf{68.13}        & \textbf{36.17}      & \textbf{39.13}    & \textbf{43.90}      & 88.66          \\
 % + SFT (R5 Data w/o CoT)     & 16.66 & 11.29  & 2.55   & 0.55  & 63.13        & 31.56      & 37.16    & 42.75      & \textbf{9.73}           \\
\midrule
\multicolumn{4}{l}{\textit{Results on DeepSeek-R1-Distilled-Qwen3-8B:}} \\
Raw Model                             & 8.53  & 4.42   & 2.16   & 0.00  & 48.75        & 13.02      & 18.70                         & 29.79                      & 97.02           \\
 + SFT (Source Code)                   & 3.79  & 1.34   & 1.89   & 0.13  & 31.88        & 5.97       & 0.00                          & 24.81                      & 98.74           \\
 + SFT (R0 Data)                   & 10.66 & 7.65   & 1.70   & 0.19  & 52.50        & 23.37      & 28.40                         & 30.78                      & 98.88           \\
 + SFT (R5 Data)             & 11.86 & 7.94   & 2.18   & 0.43  & \textbf{55.00}        & \textbf{23.77}      & \textbf{28.82}                         & \textbf{65.74}                      & \textbf{99.37}           \\
\bottomrule
\end{tabular}
% \vspace{-0.3cm}
\end{table}

\textbf{Benefits of self-debugging.}
% We also observe that the self-debugging pipeline substantially enhances the LLMs’ UT-generation capability by 71\%.
Compared to SFT on data directly generated by DeepSeek-R1 without self-debugging (Round-0 data), the Round-5 data increases the pass rate by 14\% (from 31.63 to 36.17) and the line coverage by 18\% (from 33.13 to 39.13), demonstrating the substantial improvement in both correctness and completeness metrics.
These improvement indicates that higher-quality SFT data enable LLMs to better capture program semantics, leading to improved functional understanding and broader execution coverage.

\begin{tcolorbox}[colback=lightgray!25!white, colframe=gray, title=\textbf{Summary of RQ3}]
\textbf{End-to-end effectiveness.}
Our synthesized data substantially improves LLMs capability in UT generation by 45\%. 
Additionally, our self-debugging pipeline further increases the pass rate by 14\% and the line coverage by 18\%.
\end{tcolorbox}

\subsection{RQ4: To what extent do our synthesized CoTs improve LLM’s UT generation capabilities?}

Although existing work~\cite{cleantest, methods2test} shows that higher-quality test files can enhance an LLM’s test-generation capability, the specific contribution of CoTs to this task remains unclear.
To address this gap, we conduct an ablation study on the CoTs produced during our self-debugging pipeline by removing the CoTs from the SFT training data and evaluating the resulting impact on UT generation performance.

\textbf{Results of the ablation study.}
Table~\ref{tab: ablation cot} presents the evaluation results of our ablation study on the contribution of CoTs.
We observe that fine-tuning with CoTs improves both correctness metrics (execution rate and pass rate) and completeness metrics (line/branch coverage and mutation scores). 
For instance, the branch coverage of Qwen2.5-Coder-32B-Instruct decreases from 43.90 to 27.32 when CoTs are removed, demonstrating that CoTs play a critical role in guiding the model to generate more complete test cases.

% \textbf{Comparison between reasoning and non-reasoning models.}
% \tianyu{tbd after the qwen3 results.}

\begin{table}[t]
\small
\centering
\caption{The Execution Results and Evaluation Metrics of Test Files under Various Debugging Strategies}
\label{tab: ablation cot}
\begin{threeparttable}
\begin{tabular}{lrcccccccr}
\toprule
                    & \#tests & \#passed & \#failed & \#error & exec & pass & $Cov_{l}$ & $Cov_{b}$ & $S_{mut}$ \\
\midrule
\multicolumn{4}{l}{\textit{Results on Qwen2.5-Coder-32B-Instruct:}} \\
SFT (R5 Data) & 18.78 & 12.93  & 3.66   & 0.42  & \textbf{68.13}        & \textbf{36.17}      & \textbf{39.13}    & \textbf{43.90}      & \textbf{88.66}          \\
\quad - w/o  CoT & 16.66 & 11.29  & 2.55   & 0.55  & 63.13        & 31.56      & 37.16    & 27.32      & 69.43         \\
\midrule
\multicolumn{4}{l}{\textit{DeepSeek-R1-Distilled-Qwen3-8B:}} \\
 + SFT (R5 Data)             & 11.86 & 7.94   & 2.18   & 0.43  & 55.00        & \textbf{23.77}      & \textbf{28.82}                         & \textbf{34.26}                      & \textbf{99.37}           \\
\quad - w/o  CoT     & 10.50 & 6.98   & 2.15   & 0.25  & \textbf{58.75}        & 23.49      & 26.10                         & 32.36                      & 99.16           \\
\bottomrule
\end{tabular}
% \begin{tablenotes}
%     \item Generate (R5-CoT) indicates that using Round 5's Chain-of-Thought as generation prefix to generate only test files, i.e., Q+T -> A.
% \end{tablenotes}
\end{threeparttable}
% \vspace{-0.3cm}
\end{table}

\textbf{A case study of self-debugging on CoTs.}
Given that debugging and refinement of test cases are widely adopted in existing work~\cite{coderm, cleantest}, we further examine how CoTs evolve during the self-debugging process.
Building on the motivating example in Section~\ref{sec: background}, Listing~\ref{lst: coverage debugging} illustrates how the corresponding CoT is iteratively self-debugged and subsequently compressed.

\begin{figure}
    \centering    
\begin{lstlisting}[caption = A Case Study on CoT's Self-debugging and Compression, label = lst: coverage debugging, style=cot, linebackgroundcolor={%
    \ifnum\value{lstnumber}=17\color{Red!90!white}\fi%
    \ifnum\value{lstnumber}=19\color{Green!90!white}\fi%
    \ifnum\value{lstnumber}>24\color{Green!90!white}\fi%
    \ifnum\value{lstnumber}=18\color{Green!90!white}\fi%\fi%
}]
(*@\textbf{Debugging CoT}@*):
    To cover this, we need to create an image with an affine matrix that has non-diagonal elements and requires axis swapping. Specifically, we want an affine where the axis numbers (the index of the maximum absolute value in each column) are not in increasing order.
    ...
    Then we swap the first two axes in the affine by permuting the first two rows and columns? Actually, we want to create a situation where the `axis_numbers` (the index of the max absolute value in each column) are not sorted.
    We can do: ``` affine = np.array([[0, 2, 0, 0], [3, 0, 0, 0], [0, 0, 4, 0], [0, 0, 0, 1]]) ```
    Then we create an image with this affine and data of shape (2,3,4). We call `reorder_img` with `resample=None` ...
--------------------------------------------------------------------------------------------------------------
(*@\textbf{CoT Compression}@*):
  We'll cover:
    ...
    6. resample_to_img: test resampling to a target image.
  - 7. reorder_img: test reordering the axes of an image (with and without resampling).
  + 7. reorder_img: test reordering the axes of an image (with and without resampling) and including a case
        that triggers axis swapping due to non-sorted axis numbers.
    8. BoundingBoxError: test that it is raised when expected.
    ...
For `reorder_img`:
   1. Without resampling: we flip the x-axis and reorder without resampling, then check the affine has positive diagonal.
   2. With resampling: we apply a rotation and then reorder with resampling, then check the affine has positive diagonal.
 + 3. Axis swapping: we create an affine that maps the x-axis to the y-direction and y-axis to the x-direction
    (with a non-sorted axis_numbers [1,0,2]) to cover the code that swaps axes. We then check that after 
    reordering, the affine becomes diagonal with positive entries and the axes are in standard order.
    \end{lstlisting}
% \vspace{-0.2cm}
% \caption*{
% say something
% }
\end{figure}

The CoT debugging process is shown in the \textbf{Debugging CoT} block. 
The model first reflects on the self-debugging prompt, which indicates that ``the line return header, rows in the read\_csv function is not covered,'' and then analyzes the underlying reason for this coverage gap.
Based on this analysis, the model determines that an additional test case should be added to handle an input table without body rows.

For the CoT compression step, we present the diff between the compressed CoT and the CoT from the previous round (prior to debugging and compression) in the \textbf{CoT Compression} block. 
This comparison shows that the lengthy debugging CoT is reduced to a concise version containing only one inserted line and one modified line, preserving exactly the essential information conveyed by the debugging step. 
The extended intermediate reasoning is therefore removed without loss of the critical debugging outcome.

\begin{tcolorbox}[colback=lightgray!25!white, colframe=gray, title=\textbf{Summary of RQ4}]
\textbf{Effectiveness improvement:}
CoTs provide structured reasoning that substantially enhances UT generation, as removing them leads to noticeable drops in correctness and completeness metrics.

\textbf{CoT Compression's benefits:}
Self-debugged and compressed CoTs preserve essential debugging insights while eliminating unnecessary reasoning, improving data efficiency without sacrificing quality.
\end{tcolorbox}

%% file: discuss.tex
\section{Discussion}~\label{sec: discuss}
% In this section, we first explain why our self-debugging procedure is applied only during data synthesis rather than during the fine-tuning stage. 
% We then discuss the generalizability of our \approach{} approach.

In this section, we first explain the influence of focal-file correctness during data synthesis. 
We then discuss the generalizability of our \approach{} approach.

% \subsection{Rationale for Applying Self-Debugging Only During Data Synthesis}
% As shown in Section~\ref{sec: eval fine tune}, we do not apply self-debugging when evaluating the fine-tuned LLMs, as in a real-world UT generation task, an LLM cannot assume the correctness of the focal file.
% % The primary reason is to remain consistent with real-world usage scenarios. 
% % When users query an LLM to generate a test file for a given focal file, they typically cannot verify the correctness of that focal file. 
% Consequently, if the generated test case fails, it is impossible to determine whether the error lies in the focal file, the generated test file, or both.
% \hzy{What's the relationship between the upper and following sentance}
% To avoid this ambiguity, we instruct the fine-tuned LLM to generate only a single test case and evaluate performance based on that output.

\subsection{The Influence of Focal-file Correctness During Data Synthesis}
During data synthesis, we take the focal files' behaviour as the ground-truth behaviour for self-debugging.
% In contrast, during data synthesis, 
This is because the focal files are sourced from mature open-source repositories, where their correctness has generally been validated by the original developers.
Existing work~\cite{opencodereasoning2} further demonstrates that even if some training data contain defects, they can still be effectively used in improving an LLM’s capability despite occasional imperfections in these data.
This is largely because such imperfect data often correspond to more challenging tasks, from which LLMs can learn more effectively.

To further validate this conclusion regarding UT generation, we conduct an additional ablation study on a filtered subset of \dataset{}. 
Specifically, we filter the synthesized dataset using the following criteria: (1) the test file is successfully executed (execution rate = 1); (2) the pass rate exceeds 0.3; and (3) the line coverage exceeds 0.3.
We then fine-tune Qwen2.5-Coder-32B-Instruct to assess whether this ``correct'' subset improves end-to-end effectiveness.

Table~\ref{tab: ablation quality filtering} presents the dataset statistics before and after filtering, along with the evaluation results on TestGenEval.
From these results, we observe two key findings.
(1) Although filtering substantially improves the quality of test files in the training set, it reduces both the size and diversity of the available data.
(2) Fine-tuning on the filtered subset does not yield a significant improvement in end-to-end effectiveness.
We additionally conduct a paired t-test~\cite{sedgwick2012pearson} on all evaluation metics. 
We show that only the results in mutation scores are significant, indicating that there is no significant difference between our filtered and raw dataset.

\begin{table}[t]
\small
\centering
\caption{The Evaluation Metrics of Test Files with/without Quality Filtering}
\label{tab: ablation quality filtering}
\begin{threeparttable}
\begin{tabular}{lrcccccccr}
\toprule
                    & total & passed & failed & error & exec & pass & $Cov_{l}$ & $Cov_{b}$ & $S_{mut}$ \\
\midrule
\multicolumn{4}{l}{\textit{Scores of \dataset{}'s Test Files:}} \\
w/o Filtering (74,518 tasks)         & 12.60 & 9.96   & 2.09   & 0.42  & 73.17        & 56.60      & 66.75          & 52.90            & 91.90           \\
w/ Filtering (45,598 tasks) & 17.73 & 15.41  & 2.22   & 0.04  & \textbf{100.00}       & \textbf{87.58}      & \textbf{74.81}    & \textbf{62.63}            & \textbf{94.52}           \\
\midrule
\multicolumn{5}{l}{\textit{Results on Fine-Tuning Qwen2.5-Coder-32B-Instruct:}} \\
w/o Filtering & 18.78 & 12.93  & 3.66   & 0.42  & 68.13        & \textbf{36.17}      & 39.13    & \textbf{43.90}      & 88.66          \\
w/ Filtering   & 21.85 & 14.38  & 4.47   & 0.21  & \textbf{71.25}        & 34.24      & \textbf{41.60}    & 43.66      & \textbf{96.90}           \\
- \textit{p value}  & - & -  & -   & -  & 0.517        & 0.549      & 0.284    & 0.196      & \textit{0.005}           \\
\bottomrule
\end{tabular}
\begin{tablenotes}
    \item We highlight all significant P-values ($<0.05$) in Italic.
\end{tablenotes}
\end{threeparttable}
\end{table}

% spec to test/ intension to test

% programming languages

\subsection{Generalizability}
\approach{} is designed with strong generalization capability and is not limited to Python repositories.
For example, when applied to UT generation in Java repositories, method2test~\cite{methods2test} can be used to generate initial tests, and self-debugging can be conducted in the same pipeline.

Moreover, our self-debugging pipeline does not rely on ground-truth test files, which are only used for evaluation. 
As a result, the pipeline can be incorporated into general data synthesis tasks that involve third-party feedback, e.g., execution results of generated test cases. 
We believe that this pipeline demonstrates a novel application of self-evolving and expert-iteration principles to data synthesis tasks, especially when combined with Chain-of-Thought reasoning.

%% file: threats.tex
\section{Threats to Validity} \label{sec: threats}

A major threat to external validity arises from the generalizability of our source dataset, pymethod2tes.
Although the dataset encompasses a wide range of Python repositories, it may still omit certain categories of projects. For instance, pymethod2test does not include the most recent versions of deep-learning frameworks such as PyTorch and TensorFlow, which evolve rapidly.
Moreover, these frameworks typically exhibit poor version compatibility, further complicating LLM generalization across different releases.
To mitigate this threat, we removed any overlapping repositories between our training set (pymethod2test) and testing set (TestGenEval), enabling a more accurate assessment of the generalization ability of our synthesized dataset. 
As shown in Section~\ref{sec: eval fine tune}, the UT generation capability of our fine-tuned LLMs generalizes well to these domain-specific repositories, such as SymPy (a scientific computation library) in TestGenEval.

The threats to internal validity are instrumentation effects that can bias our results.
To reduce this risk, we manually inspected intermediate artifacts—including Chains-of-Thought and generated test files—for dozens of sampled focal files, such as those shown in Listing~\ref{lst: coverage debugging}.

%% file: related.tex
\section{Related Work}

\subsection{Unit Test Generation}
A variety of techniques have been developed for automated unit test generation. In the literature, traditional test generation approaches are commonly categorized into three groups: random-based methods~\cite{papineni2002bleu}, constraint-based methods~\cite{ernst2007daikon, xiao2013characteristic, csallner2008dysy}, and search-based methods~\cite{harman2009theoretical, blasi2022call, delgado2022interevo}. 
Although these approaches have achieved promising results, they often suffer from path explosion in symbolic execution or fail to adequately explore large input spaces, limiting their ability to uncover complex bugs~\cite{papineni2002bleu}. 
Moreover, they typically face challenges related to explainability and readability~\cite{watson2020learning}.

Recently, Large Language Models (LLMs) have demonstrated strong performance across a broad range of code-related tasks~\cite{fried2022incoder, geng2024large, xia2023automated, hu2018deep, li2022codereviewer}. 
Building on this progress, several studies have explored using LLMs for test case generation~\cite{alagarsamy2024a3test, tufano2020unit, lemieux2023codamosa}. 
For example, Junwei et al.~\cite{cleantest} filter incorrect test cases from training sets to improve the capability of fine-tuned LLMs in UT generation.
Alagarsamy et al.~\cite{alagarsamy2024a3test} pre-train and fine-tune an LLM using focal methods and assertion statements to synthesize test cases. 
Similarly, Tufano et al.~\cite{tufano2020unit} pre-train an LLM on a large Java corpus and fine-tuned it using a translation task to generate unit tests.
% Hashtroudi et al.~\cite{shin2024domain} fine-tune the LLM with existing developer-written tests to generate domain adaptation tests. 
Konstantinos et al.~\cite{blast} combine search-based testing with LLMs to produce high-quality issue-reproducing tests.
Chen et al.~\cite{clast} focus on prompt-level optimization, applying post-processing techniques to enhance the clarity and effectiveness of in-context examples.

Our work differs from prior work in two main aspects.
First, existing work primarily focuses on collecting and cleaning data from open-source repositories, which inevitably introduces a trade-off between data quality and dataset scale. 
In contrast, our \approach{} method employs self-debugging to synthesize high-quality test data while preserving dataset scale, thereby offering greater potential benefits for model training.
Second, in contrast to the conclusions of previous studies, we demonstrate that for large language models, even incorrect data can still contribute to the final capability of unit test generation.

\subsection{Self-Debugging}
Self-debugging, (also including self-evolving) style of learning has been explored in many contexts.
A reinforced self-training frame work is proposed in the machine translation task~\cite{gulcehre2023reinforced}, and the expert iteration strategy is leveraged for math proving with Lean~\cite{polu2022formal}. 
Self-debugging is also adopted in code generation tasks.
For example, IterPref~\cite{wu2025iterpref} adopts self-debugging in code generation, e.g., tasks in HumanEval~\cite{codex} and LiveCodeBench~\cite{livecodebench}, and SAFE~\cite{safe} adopts self-debugging in generating formal verification.

%% file: conclusion.tex
\section{Conclusion}\label{sec: conclusion}
In this paper, we have presented \approach{}, a new data synthesis approach for improving LLM-based unit test generation by producing high-quality <focal method, test, CoT> training examples.
The pipeline combines CoT-involved self-debugging (error-, failure-, and coverage-focused iterations) with CoT compression to produce compact, faithful reasoning paired with corrected tests. 
Our evaluation results show that we have synthesized a dataset of 74,518 examples and enabled supervised fine-tuning that improves empirical test-generation results: a pass rate of 36.17\% on test assertions, a branch coverage of 43.90\%, and a mutation score of 88.66\%.

% \section{Data Availability}
% We open-source our artifacts on our anonymous website~\footnote{\url{https://github.com/SigmaDataAna/repo_smith}}.

% We also list our scan results in RQ 5 to show our effectiveness in real-world security practice.
% We do not open-source our code repository now due to the security requirement of our industry partner, and we plan to open-source it at the publication of our paper.
% However, in Section~\ref{sec: approach}, we list all details of our approach, so that our approach can be easily reproduced.

% \section{Data Availability}
% We list 225 <vulnerability, library> pairs to show high value of security practice on our anonymous website~\cite{repo}.
% We plan to open-source our  \javadata{} dataset, and all evaluation results at the publication of our paper. 
% We do not open-source our code repository now due to the security requirement of our industry partner.
% However, in Section~\ref{sec:approach}, we list all details of our approach, so that our approach can be easily reproduced.

%% file: data_availability.tex
\section{Data Availability}
To facilitate reproducibility and further research, we make our implementation publicly available. The code repository, Repo-Smith, is released on an anonymous GitHub repository at: \url{https://anonymous.4open.science/r/Repo-smith-EBE5}.

% The datasets used in this work are publicly accessible. PyMethod2Test is available via Zenodo at \url{https://zenodo.org/records/14264519}
% , and TestGenEval can be obtained from Hugging Face at \url{https://huggingface.co/datasets/kjain14/testgeneval}
% .
% In addition, the synthesized dataset using self-debugging is not publicly released at the time of submission, but can be reproduced using \approach{} to perform self-debugging on PyMethod2Test. 
% We do not open-source \dataset{} due to the security requirements of our industrial partner, and we plan to open-source \dataset{} in the future. 

%% file: main.bib
@inproceedings{reanu2008combining,
  title={Combining unit-level symbolic execution and system-level concrete execution for testing NASA software},
  author={Pǎsǎreanu, Corina S and Mehlitz, Peter C and Bushnell, David H and Gundy-Burlet, Karen and Lowry, Michael and Person, Suzette and Pape, Mark},
  booktitle={Proceedings of the 2008 international symposium on Software testing and analysis},
  pages={15--26},
  year={2008}
}

@inproceedings{xie2005symstra,
  title={Symstra: A framework for generating object-oriented unit tests using symbolic execution},
  author={Xie, Tao and Marinov, Darko and Schulte, Wolfram and Notkin, David},
  booktitle={International Conference on Tools and Algorithms for the Construction and Analysis of Systems},
  pages={365--381},
  year={2005},
  organization={Springer}
}

@inproceedings{evosuite,
  title={Evosuite: automatic test suite generation for object-oriented software},
  author={Fraser, Gordon and Arcuri, Andrea},
  booktitle={Proceedings of the 19th ACM SIGSOFT symposium and the 13th European conference on Foundations of software engineering},
  pages={416--419},
  year={2011}
}

@article{gargantini1999using,
  title={Using model checking to generate tests from requirements specifications},
  author={Gargantini, Angelo and Heitmeyer, Constance},
  journal={ACM SIGSOFT Software Engineering Notes},
  volume={24},
  number={6},
  pages={146--162},
  year={1999},
  publisher={ACM New York, NY, USA}
}

@article{enoiu2016automated,
  title={Automated test generation using model checking: an industrial evaluation},
  author={Enoiu, Eduard P and {\v{C}}au{\v{s}}evi{\'c}, Adnan and Ostrand, Thomas J and Weyuker, Elaine J and Sundmark, Daniel and Pettersson, Paul},
  journal={International Journal on Software Tools for Technology Transfer},
  volume={18},
  number={3},
  pages={335--353},
  year={2016},
  publisher={Springer}
}

@article{cleantest,
  title={Less Is More: On the Importance of Data Quality for Unit Test Generation},
  author={Zhang, Junwei and Hu, Xing and Gao, Shan and Xia, Xin and Lo, David and Li, Shanping},
  journal={Proceedings of the ACM on Software Engineering},
  volume={2},
  number={FSE},
  pages={1293--1316},
  year={2025},
  publisher={ACM New York, NY, USA}
}

@inproceedings{methods2test,
  title={Methods2Test: A dataset of focal methods mapped to test cases},
  author={Tufano, Michele and Deng, Shao Kun and Sundaresan, Neel and Svyatkovskiy, Alexey},
  booktitle={Proceedings of the 19th International Conference on Mining Software Repositories},
  pages={299--303},
  year={2022}
}

@inproceedings{pymethods2test,
  title={pyMethods2Test: A dataset of Python tests mapped to focal methods},
  author={Abdelmadjid, Idriss and Dyer, Robert},
  booktitle={2025 IEEE/ACM 22nd International Conference on Mining Software Repositories (MSR)},
  pages={846--850},
  year={2025},
  organization={IEEE}
}

@article{yang2024empirical,
  title={An empirical study of unit test generation with large language models},
  author={Yang, Lin and Yang, Chen and Gao, Shutao and Wang, Weijing and Wang, Bo and Zhu, Qihao and Chu, Xiao and Zhou, Jianyi and Liang, Guangtai and Wang, Qianxiang and others},
  journal={arXiv preprint arXiv:2406.18181},
  year={2024}
}

@article{coderm,
  title={Dynamic scaling of unit tests for code reward modeling},
  author={Ma, Zeyao and Zhang, Xiaokang and Zhang, Jing and Yu, Jifan and Luo, Sijia and Tang, Jie},
  journal={arXiv preprint arXiv:2501.01054},
  year={2025}
}

@inproceedings{hits,
  title={Hits: High-coverage llm-based unit test generation via method slicing},
  author={Wang, Zejun and Liu, Kaibo and Li, Ge and Jin, Zhi},
  booktitle={Proceedings of the 39th IEEE/ACM International Conference on Automated Software Engineering},
  pages={1258--1268},
  year={2024}
}

@article{wu2025iterpref,
  title={Iterpref: Focal preference learning for code generation via iterative debugging},
  author={Wu, Jie and Li, Haoling and Zhang, Xin and Luo, Jianwen and Huang, Yangyu and Chu, Ruihang and Yang, Yujiu and Li, Scarlett},
  journal={arXiv preprint arXiv:2503.02783},
  year={2025}
}

@inproceedings{safe,
  title={Automated Proof Generation for Rust Code via Self-Evolution},
  author={Chen, Tianyu and Lu, Shuai and Lu, Shan and Gong, Yeyun and Yang, Chenyuan and Li, Xuheng and Misu, Md Rakib Hossain and Yu, Hao and Duan, Nan and CHENG, Peng and others},
  booktitle={The Thirteenth International Conference on Learning Representations},
  year={2025}
}

@article{opencodereasoning,
  title={Opencodereasoning: Advancing data distillation for competitive coding},
  author={Ahmad, Wasi Uddin and Narenthiran, Sean and Majumdar, Somshubra and Ficek, Aleksander and Jain, Siddhartha and Huang, Jocelyn and Noroozi, Vahid and Ginsburg, Boris},
  journal={arXiv preprint arXiv:2504.01943},
  year={2025}
}

@article{opencodereasoning2,
  title={OpenCodeReasoning-II: A Simple Test Time Scaling Approach via Self-Critique},
  author={Ahmad, Wasi Uddin and Majumdar, Somshubra and Ficek, Aleksander and Narenthiran, Sean and Samadi, Mehrzad and Huang, Jocelyn and Jain, Siddhartha and Noroozi, Vahid and Ginsburg, Boris},
  journal={arXiv preprint arXiv:2507.09075},
  year={2025}
}

@article{codex,
  title={Evaluating large language models trained on code},
  author={Chen, Mark and Tworek, Jerry and Jun, Heewoo and Yuan, Qiming and Pinto, Henrique Ponde De Oliveira and Kaplan, Jared and Edwards, Harri and Burda, Yuri and Joseph, Nicholas and Brockman, Greg and others},
  journal={arXiv preprint arXiv:2107.03374},
  year={2021}
}

@article{livecodebench,
  title={Livecodebench: Holistic and contamination free evaluation of large language models for code},
  author={Jain, Naman and Han, King and Gu, Alex and Li, Wen-Ding and Yan, Fanjia and Zhang, Tianjun and Wang, Sida and Solar-Lezama, Armando and Sen, Koushik and Stoica, Ion},
  journal={arXiv preprint arXiv:2403.07974},
  year={2024}
}

@article{testgeneval,
  title={Testgeneval: A real world unit test generation and test completion benchmark},
  author={Jain, Kush and Synnaeve, Gabriel and Roziere, Baptiste},
  journal={arXiv preprint arXiv:2410.00752},
  year={2024}
}

@inproceedings{testeval,
  title={Testeval: Benchmarking large language models for test case generation},
  author={Wang, Wenhan and Yang, Chenyuan and Wang, Zhijie and Huang, Yuheng and Chu, Zhaoyang and Song, Da and Zhang, Lingming and Chen, An Ran and Ma, Lei},
  booktitle={Findings of the Association for Computational Linguistics: NAACL 2025},
  pages={3547--3562},
  year={2025}
}

@article{swebench,
  title={Swe-bench: Can language models resolve real-world github issues?},
  author={Jimenez, Carlos E and Yang, John and Wettig, Alexander and Yao, Shunyu and Pei, Kexin and Press, Ofir and Narasimhan, Karthik},
  journal={arXiv preprint arXiv:2310.06770},
  year={2023}
}

@inproceedings{papineni2002bleu,
  title={Bleu: a method for automatic evaluation of machine translation},
  author={Papineni, Kishore and Roukos, Salim and Ward, Todd and Zhu, Wei-Jing},
  booktitle={Proceedings of the 40th annual meeting of the Association for Computational Linguistics},
  pages={311--318},
  year={2002}
}

@inproceedings{csallner2008dysy,
  title={DySy: Dynamic symbolic execution for invariant inference},
  author={Csallner, Christoph and Tillmann, Nikolai and Smaragdakis, Yannis},
  booktitle={Proceedings of the 30th international conference on Software engineering},
  pages={281--290},
  year={2008}
}

@article{ernst2007daikon,
  title={The Daikon system for dynamic detection of likely invariants},
  author={Ernst, Michael D and Perkins, Jeff H and Guo, Philip J and McCamant, Stephen and Pacheco, Carlos and Tschantz, Matthew S and Xiao, Chen},
  journal={Science of computer programming},
  volume={69},
  number={1-3},
  pages={35--45},
  year={2007},
  publisher={Elsevier}
}

@inproceedings{xiao2013characteristic,
  title={Characteristic studies of loop problems for structural test generation via symbolic execution},
  author={Xiao, Xusheng and Li, Sihan and Xie, Tao and Tillmann, Nikolai},
  booktitle={2013 28th IEEE/ACM International Conference on Automated Software Engineering (ASE)},
  pages={246--256},
  year={2013},
  organization={IEEE}
}

@inproceedings{blasi2022call,
  title={Call me maybe: Using nlp to automatically generate unit test cases respecting temporal constraints},
  author={Blasi, Arianna and Gorla, Alessandra and Ernst, Michael D and Pezz{\`e}, Mauro},
  booktitle={Proceedings of the 37th IEEE/ACM International Conference on Automated Software Engineering},
  pages={1--11},
  year={2022}
}

@article{delgado2022interevo,
  title={InterEvo-TR: Interactive evolutionary test generation with readability assessment},
  author={Delgado-P{\'e}rez, Pedro and Ram{\'\i}rez, Aurora and Valle-G{\'o}mez, Kevin J and Medina-Bulo, Inmaculada and Romero, Jos{\'e} Ra{\'u}l},
  journal={IEEE Transactions on Software Engineering},
  volume={49},
  number={4},
  pages={2580--2596},
  year={2022},
  publisher={IEEE}
}

@article{harman2009theoretical,
  title={A theoretical and empirical study of search-based testing: Local, global, and hybrid search},
  author={Harman, Mark and McMinn, Phil},
  journal={IEEE Transactions on Software Engineering},
  volume={36},
  number={2},
  pages={226--247},
  year={2009},
  publisher={IEEE}
}

@article{seed2025seed,
  title={Seed-coder: Let the code model curate data for itself},
  author={Seed, ByteDance and Zhang, Yuyu and Su, Jing and Sun, Yifan and Xi, Chenguang and Xiao, Xia and Zheng, Shen and Zhang, Anxiang and Liu, Kaibo and Zan, Daoguang and others},
  journal={arXiv preprint arXiv:2506.03524},
  year={2025}
}

@article{deepseekr1,
  title={Deepseek-r1: Incentivizing reasoning capability in llms via reinforcement learning},
  author={Guo, Daya and Yang, Dejian and Zhang, Haowei and Song, Junxiao and Zhang, Ruoyu and Xu, Runxin and Zhu, Qihao and Ma, Shirong and Wang, Peiyi and Bi, Xiao and others},
  journal={arXiv preprint arXiv:2501.12948},
  year={2025}
}

@Misc{o4-mini,
author = {OpenAI},
howpublished = "\url{https://openai.com/index/introducing-o3-and-o4-mini/}",
note = {},
title = {o4-mini},
year = {2026}
}

@inproceedings{lahiri2024evaluating,
  title={Evaluating llm-driven user-intent formalization for verification-aware languages},
  author={Lahiri, Shuvendu K},
  booktitle={CONFERENCE ON FORMAL METHODS IN COMPUTER-AIDED DESIGN--FMCAD 2024},
  pages={142},
  year={2024}
}

@article{baker2012empirical,
  title={An empirical evaluation of mutation testing for improving the test quality of safety-critical software},
  author={Baker, Richard and Habli, Ibrahim},
  journal={IEEE Transactions on Software Engineering},
  volume={39},
  number={6},
  pages={787--805},
  year={2012},
  publisher={IEEE}
}

@inproceedings{kaufman2022prioritizing,
  title={Prioritizing mutants to guide mutation testing},
  author={Kaufman, Samuel J and Featherman, Ryan and Alvin, Justin and Kurtz, Bob and Ammann, Paul and Just, Ren{\'e}},
  booktitle={Proceedings of the 44th International Conference on Software Engineering},
  pages={1743--1754},
  year={2022}
}

@inproceedings{whalen2006coverage,
  title={Coverage metrics for requirements-based testing},
  author={Whalen, Michael W and Rajan, Ajitha and Heimdahl, Mats PE and Miller, Steven P},
  booktitle={Proceedings of the 2006 international symposium on Software testing and analysis},
  pages={25--36},
  year={2006}
}

@article{aghababaeyan2023black,
  title={Black-box testing of deep neural networks through test case diversity},
  author={Aghababaeyan, Zohreh and Abdellatif, Manel and Briand, Lionel and Bagherzadeh, Mojtaba and others},
  journal={IEEE Transactions on Software Engineering},
  volume={49},
  number={5},
  pages={3182--3204},
  year={2023},
  publisher={IEEE}
}

@article{qwen25coder,
  title={Qwen2.5-coder technical report},
  author={Hui, Binyuan and Yang, Jian and Cui, Zeyu and Yang, Jiaxi and Liu, Dayiheng and Zhang, Lei and Liu, Tianyu and Zhang, Jiajun and Yu, Bowen and Lu, Keming and others},
  journal={arXiv preprint arXiv:2409.12186},
  year={2024}
}

@inproceedings{watson2020learning,
  title={On learning meaningful assert statements for unit test cases},
  author={Watson, Cody and Tufano, Michele and Moran, Kevin and Bavota, Gabriele and Poshyvanyk, Denys},
  booktitle={Proceedings of the ACM/IEEE 42nd International Conference on Software Engineering},
  pages={1398--1409},
  year={2020}
}

@inproceedings{xia2023automated,
  title={Automated program repair in the era of large pre-trained language models},
  author={Xia, Chunqiu Steven and Wei, Yuxiang and Zhang, Lingming},
  booktitle={2023 IEEE/ACM 45th International Conference on Software Engineering (ICSE)},
  pages={1482--1494},
  year={2023},
  organization={IEEE}
}

@article{fried2022incoder,
  title={Incoder: A generative model for code infilling and synthesis},
  author={Fried, Daniel and Aghajanyan, Armen and Lin, Jessy and Wang, Sida and Wallace, Eric and Shi, Freda and Zhong, Ruiqi and Yih, Wen-tau and Zettlemoyer, Luke and Lewis, Mike},
  journal={arXiv preprint arXiv:2204.05999},
  year={2022}
}

@inproceedings{geng2024large,
  title={Large language models are few-shot summarizers: Multi-intent comment generation via in-context learning},
  author={Geng, Mingyang and Wang, Shangwen and Dong, Dezun and Wang, Haotian and Li, Ge and Jin, Zhi and Mao, Xiaoguang and Liao, Xiangke},
  booktitle={Proceedings of the 46th IEEE/ACM International Conference on Software Engineering},
  pages={1--13},
  year={2024}
}

@inproceedings{hu2018deep,
  title={Deep code comment generation},
  author={Hu, Xing and Li, Ge and Xia, Xin and Lo, David and Jin, Zhi},
  booktitle={Proceedings of the 26th conference on program comprehension},
  pages={200--210},
  year={2018}
}

@article{li2022codereviewer,
  title={Codereviewer: Pre-training for automating code review activities},
  author={Li, Zhiyu and Lu, Shuai and Guo, Daya and Duan, Nan and Jannu, Shailesh and Jenks, Grant and Majumder, Deep and Green, Jared and Svyatkovskiy, Alexey and Fu, Shengyu and others},
  journal={arXiv preprint arXiv:2203.09095},
  year={2022}
}

@article{alagarsamy2024a3test,
  title={A3test: Assertion-augmented automated test case generation},
  author={Alagarsamy, Saranya and Tantithamthavorn, Chakkrit and Aleti, Aldeida},
  journal={Information and Software Technology},
  volume={176},
  pages={107565},
  year={2024},
  publisher={Elsevier}
}

@inproceedings{lemieux2023codamosa,
  title={Codamosa: Escaping coverage plateaus in test generation with pre-trained large language models},
  author={Lemieux, Caroline and Inala, Jeevana Priya and Lahiri, Shuvendu K and Sen, Siddhartha},
  booktitle={2023 IEEE/ACM 45th International Conference on Software Engineering (ICSE)},
  pages={919--931},
  year={2023},
  organization={IEEE}
}

@article{tufano2020unit,
  title={Unit test case generation with transformers and focal context},
  author={Tufano, Michele and Drain, Dawn and Svyatkovskiy, Alexey and Deng, Shao Kun and Sundaresan, Neel},
  journal={arXiv preprint arXiv:2009.05617},
  year={2020}
}

@article{sedgwick2012pearson,
  title={Pearson’s correlation coefficient},
  author={Sedgwick, Philip},
  journal={Bmj},
  volume={345},
  year={2012},
  publisher={British Medical Journal Publishing Group}
}

@article{zhu2025uncertainty,
  title={Uncertainty-guided chain-of-thought for code generation with llms},
  author={Zhu, Yuqi and Li, Ge and Jiang, Xue and Li, Jia and Mei, Hong and Jin, Zhi and Dong, Yihong},
  journal={arXiv preprint arXiv:2503.15341},
  year={2025}
}

@article{yang2024chain,
  title={Chain-of-thought in neural code generation: From and for lightweight language models},
  author={Yang, Guang and Zhou, Yu and Chen, Xiang and Zhang, Xiangyu and Zhuo, Terry Yue and Chen, Taolue},
  journal={IEEE Transactions on Software Engineering},
  year={2024},
  publisher={IEEE}
}

@article{phi15,
  title={Textbooks are all you need ii: phi-1.5 technical report},
  author={Li, Yuanzhi and Bubeck, S{\'e}bastien and Eldan, Ronen and Del Giorno, Allie and Gunasekar, Suriya and Lee, Yin Tat},
  journal={arXiv preprint arXiv:2309.05463},
  year={2023}
}

@article{phi,
  title={Textbooks are all you need},
  author={Gunasekar, Suriya and Zhang, Yi and Aneja, Jyoti and Mendes, Caio C{\'e}sar Teodoro and Del Giorno, Allie and Gopi, Sivakanth and Javaheripi, Mojan and Kauffmann, Piero and de Rosa, Gustavo and Saarikivi, Olli and others},
  journal={arXiv preprint arXiv:2306.11644},
  year={2023}
}

@article{laban2025llms,
  title={Llms get lost in multi-turn conversation},
  author={Laban, Philippe and Hayashi, Hiroaki and Zhou, Yingbo and Neville, Jennifer},
  journal={arXiv preprint arXiv:2505.06120},
  year={2025}
}

@article{zhang2024imperative,
  title={The imperative of conversation analysis in the era of llms: A survey of tasks, techniques, and trends},
  author={Zhang, Xinghua and Yu, Haiyang and Li, Yongbin and Wang, Minzheng and Chen, Longze and Huang, Fei},
  journal={arXiv preprint arXiv:2409.14195},
  year={2024}
}

@inproceedings{tsakpinis2024analyzing,
  title={Analyzing the accessibility of github repositories for pypi and npm libraries},
  author={Tsakpinis, Alexandros and Pretschner, Alexander},
  booktitle={Proceedings of the 28th International Conference on Evaluation and Assessment in Software Engineering},
  pages={345--350},
  year={2024}
}

@article{gulcehre2023reinforced,
  title={Reinforced self-training (rest) for language modeling},
  author={Gulcehre, Caglar and Paine, Tom Le and Srinivasan, Srivatsan and Konyushkova, Ksenia and Weerts, Lotte and Sharma, Abhishek and Siddhant, Aditya and Ahern, Alex and Wang, Miaosen and Gu, Chenjie and others},
  journal={arXiv preprint arXiv:2308.08998},
  year={2023}
}

@article{polu2022formal,
  title={Formal mathematics statement curriculum learning},
  author={Polu, Stanislas and Han, Jesse Michael and Zheng, Kunhao and Baksys, Mantas and Babuschkin, Igor and Sutskever, Ilya},
  journal={arXiv preprint arXiv:2202.01344},
  year={2022}
}

@article{focalstudy,
  title={An Empirical Study on Focal Methods in Deep-Learning-Based Approaches for Assertion Generation},
  author={He, Yibo and Huang, Jiaming and Yu, Hao and Xie, Tao},
  journal={Proceedings of the ACM on Software Engineering},
  volume={1},
  number={FSE},
  pages={1750--1771},
  year={2024},
  publisher={ACM New York, NY, USA}
}

@article{clast,
  title={Clarifying Semantics of In-Context Examples for Unit Test Generation},
  author={Yang, Chen and Yang, Lin and Wang, Ziqi and Wang, Dong and Zhou, Jianyi and Chen, Junjie},
  journal={arXiv preprint arXiv:2510.01994},
  year={2025}
}

@article{blast,
  title={Automated generation of issue-reproducing tests by combining llms and search-based testing},
  author={Kitsios, Konstantinos and Castelluccio, Marco and Bacchelli, Alberto},
  journal={arXiv preprint arXiv:2509.01616},
  year={2025}
}
